%% file: main.tex
\documentclass{article}
\usepackage{iclr2027_conference,times}
\iclrfinalcopy 
\usepackage{amsmath,amssymb}
\usepackage{graphicx}
\usepackage{booktabs,array,longtable}
\usepackage{placeins}
\usepackage{needspace}
\usepackage{tikz}
\usepackage{url}
\usepackage[hidelinks]{hyperref}
\usetikzlibrary{positioning,arrows.meta}
\graphicspath{{figures/}}
\newcommand{\EER}{\mathrm{EER}}
\newcommand{\EERhat}{\widehat{\mathrm{EER}}}
\newcommand{\IE}{\mathrm{IE}}
\newcommand{\NLI}{\mathrm{NLI}}
\newcommand{\mindcf}{\mathrm{minDCF}}
\newcommand{\clean}{\mathrm{clean}}

\title{InterBias-SV: Compound Conditions\\in Speaker Verification}
\author{Kamel Kamel, Hridoy Sankar Dutta, Keshav Sood, Sunil Aryal \\
School of Information Technology, Deakin University \\
Waurn Ponds, VIC, Australia \\
\texttt{kamel.kamel@research.deakin.edu.au}}
\begin{document}
\maketitle
\lhead{Preprint} 

\begin{abstract}
Speaker verification systems encounter combinations of noise, channel distortion, and changes in speech. Evaluating each condition separately does not establish whether their effects add. InterBias-SV organises this question around a four-term comparison: joint error, two marginal errors, and a common reference. Its results artefact contains 4{,}068 scored records across 17 experiments, 12 encoder labels, and six speech corpora, totalling $1.20\times10^7$ trial evaluations. Three experiment families contain the same-corpus termsneeded to compute additive contrasts. For labels assigned to speaker-trained encoders, their mean contrasts are $+0.0026$, $+0.0088$, and $+0.0024$ in equal error rate (EER), with larger variation across settings. These descriptive averages do not establish equivalence to additivity: trial matching, checkpoint identity, and parts of the condition metadata remain unverified. We also examine two interpretation problems. Near-chance EER can make additive predictions difficult to interpret, but chance performance is not a hard EER ceiling, and correlation with the prediction does not identify a saturation mechanism. Ratios of demographic gaps are unstable when their clean reference is near zero; absolute gaps provide a more direct summary. The benchmark provides condition definitions, analysis scripts, and explicit requirements for interpretable compound-condition comparisons, while separating recomputable summaries from claims that require further experimental validation.
\end{abstract}

\section{Introduction}
\label{sec:intro}
A speaker verification (SV) system may receive noisy, compressed speech from a
caller whose speaking style differs from enrolment. A benchmark can measure the
error associated with each condition, but those measurements alone do not show
how the conditions combine. The practical question is whether a joint condition
produces the error predicted by adding the two single-condition changes to a
common reference. Answering it requires measuring the joint condition and its
reference terms under comparable sampling and scoring procedures.

Existing resources establish useful but different comparisons. VoxCeleb and the
VoxCeleb Speaker Recognition Challenges evaluate recognition in unconstrained
recordings \citep{nagrani2017voxceleb,chung2018voxceleb2,nagrani2024voxsrc}.
SVeritas evaluates a broad set of stressors, including compounded noise,
reverberation, and codec conditions \citep{baali2025sveritas}. Such joint-condition
results describe difficulty; attributing a departure from additivity additionally
requires the corresponding marginals and an explicit contrast. Our focus is this
comparison, rather than a claim that prior benchmarks evaluate only isolated
stressors.

InterBias-SV combines condition definitions, an evaluation harness, and an
analysis of an existing results artefact. Its proposed evaluation unit is a
four-term comparison for one encoder and a specified factor setting. We distinguish
acoustic manipulations from population strata: an age-group contrast measures
how codec effects differ between populations, not the effect of changing a
speaker's age. We also distinguish the presence of four numerical terms from
evidence that they form a controlled experiment. That distinction matters here:
the artefact permits three families of same-corpus contrasts, but the supplied
code and metadata do not establish all the matching and provenance requirements.

\Needspace{4\baselineskip}
The contributions are:
\begin{enumerate}
\item A benchmark organisation for compound-condition SV evaluation, with
17 recorded experiments, 4{,}068 scored records, and scripts that reconstruct
its tabulated summaries (Section~\ref{sec:benchmark}).
\item Explicit requirements for interpreting additive contrasts: comparable
populations and trial sampling, measured reference terms, and a clearly defined
metric scale (Section~\ref{sec:protocol}). The accompanying audit identifies
which requirements the current artefact does and does not establish.
\item Descriptive evidence on compound-condition errors and demographic gaps
(Section~\ref{sec:results}). Small positive mean contrasts coexist with large
setting-level deviations. Near-chance performance and small reference gaps
limit interpretations of additive and ratio summaries.
\end{enumerate}
The paper contributes an evaluation framework and empirical analysis, rather
than a new encoder or a newly collected speech dataset. Its usefulness depends
on making both the measurements and their limits reproducible.

\section{Related Work and Positioning}
\label{sec:related}
\paragraph{Robustness evaluation.}
VoxCeleb provides in-the-wild speech and verification protocols
\citep{nagrani2017voxceleb,chung2018voxceleb2}; VoxSRC extends evaluation across
training and adaptation settings \citep{nagrani2024voxsrc}. Noise and reverberation
augmentation are also established components of speaker-embedding pipelines
\citep{ko2017reverberation,snyder2018xvectors}. SVeritas is the closest benchmark:
it evaluates duration, speaking conditions, noise, reverberation, channels,
codecs, age, spoofing, and adversarial perturbations, with demographic breakdowns
\citep{baali2025sveritas}. It includes compound conditions. InterBias-SV's
specific emphasis is the contrast between a joint error and the additive
prediction from its own reference terms. This is an application of a factorial
contrast, not a new statistical definition of interaction.

\paragraph{Representations and demographic evaluation.}
SUPERB evaluates speech representations across downstream tasks
\citep{yang2021superb}. WavLM, HuBERT, and UniSpeech-SAT support speaker-related
applications \citep{chen2022wavlm,hsu2021hubert,chen2022unispeechsat}, but a frozen
mean-pooled representation is a different system from one with a trained speaker
head. Our comparison retains that distinction. Work on demographic bias in
speech recognition \citep{koenecke2020racial} and speaker recognition
\citep{hutiri2022bias,fenu2021fairvoice,peri2023bias} motivates subgroup reporting.
Here we examine how reported gaps vary across conditions, without treating
subgroup EERs as a fairness assessment at a shared deployment threshold.

\begin{table}[htbp]
\centering
\caption{Evaluation emphasis of related resources. The comparison describes
reported analyses, not an exhaustive claim about what their data could support.}
\label{tab:compare}
\small
\begin{tabular}{@{}p{0.27\linewidth}p{0.68\linewidth}@{}}
\toprule Resource & Evaluation emphasis \\
\midrule
VoxCeleb / VoxSRC & In-the-wild verification and challenge protocols \\
SUPERB & Downstream evaluation of speech representations \\
Fair Voice Biometrics & Demographic imbalance and group fairness \\
SVeritas & Broad stressor coverage, including compound conditions \\
InterBias-SV & Additive contrasts and their sampling and metric requirements \\
\bottomrule
\end{tabular}
\end{table}

\section{Task, Estimand and Validity Conditions}
\label{sec:protocol}
\paragraph{Task and scoring.}
The intended task is text-independent SV. A frozen encoder maps an utterance to
an $L_2$-normalised embedding, and cosine similarity scores each genuine or
impostor trial. The artefact reports equal error rate (EER), normalised minimum
detection cost ($\mindcf$; target prior $0.01$, miss and false-accept costs both
1), and area under the receiver operating characteristic curve (AUC).
The supplied EER routine selects the ROC point with the smallest absolute
difference between false-positive and false-negative rates and averages those
rates; it does not interpolate the crossing. All EERs here are fractions, so
$0.01$ corresponds to one percentage point. No score calibration, score
normalisation, fine-tuning, or domain adaptation is specified. A common scoring
rule removes one source of variation, but does not by itself control trial
sampling or establish checkpoint identity.

\paragraph{The estimand.}
For conditions $A$ and $B$ and reference $\clean$, define
\begin{equation}
\EERhat(A\cap B)=\EER(A)+\EER(B)-\EER(\clean),\qquad
\IE(A,B)=\EER(A\cap B)-\EERhat(A\cap B).
\label{eq:ie}
\end{equation}
The non-linearity index is $\NLI(A,B)=\EER(A\cap B)/\EERhat(A\cap B)$
when the denominator exceeds $10^{-9}$; zero or negative predictions retain their
$\IE$ values but have undefined $\NLI$. Small positive denominators can still
make the ratio unstable.
Positive $\IE$ indicates error above the additive prediction on the EER scale;
negative $\IE$ indicates error below it. Neither sign alone identifies a causal
mechanism. Additivity is scale-dependent and does not imply statistical
independence of the factors. We report means and standard deviations (SDs)
over encoder--setting combinations; these SDs are not confidence intervals.

\paragraph{Requirements for interpretation.}
\label{sec:identifiability}
\textbf{C1 (comparable source population):} all terms should refer to the same
corpus and a documented target population. A shared corpus alone does not
control speaker, session, or content composition.
\textbf{C2 (comparable trials):} retain trial identities across acoustic
transformations, and match speaker and content composition when a factor
requires different recordings. Report sampling, class counts, exclusions, and
uncertainty. Independent trial draws can estimate population contrasts under an
appropriate design, but their variation must be accounted for.
\textbf{C3 (measured terms):} both marginals and the reference must be measured;
substituting the reference for a missing marginal changes the estimand.
\textbf{C4 (metric domain):} check the attainable metric range and the stability
of ratio denominators. These are requirements for the intended analysis, not
conditions certified by the presence of result rows.

Three experiment families have the same-corpus terms needed for numerical
recomputation: E13 (age group\,$\times$\,codec), E14
(emotion\,$\times$\,noise), and E17 (recorded drift label\,$\times$\,codec).
The updated E13 and E14 exports identify a common protocol cohort within each
experiment and include checkpoint hashes. They are labelled paired-age and
matched-RAVDESS runs, respectively; the aggregate CSV itself contains no trial
identities, so C2 cannot be independently checked from this release. E13 compares
an age stratum with the overlapping all-age population, rather than measuring
within-speaker ageing. E14 uses all seven non-neutral RAVDESS emotions crossed
with three noise labels and three SNRs. E17 still requires recording-level dates
that the bundled legacy loader does not provide (Section~\ref{sec:limitations}).

E15 and E18 are reported as within-corpus clean/codec summaries, not isolated
pathology or accent interactions: their reference populations belong to
different corpora. E16 (gender\,$\times$\,noise) is analysed through subgroup gaps. E20's recorded
condition values are retained descriptively. The run audit found measured
reference terms, but temporal and age proxies limit their interpretation. The analysis therefore distinguishes
recomputable four-term contrasts from verified controlled experiments.

\begin{figure}[t]
\centering
\begin{tikzpicture}[font=\small, node distance=7mm,
 box/.style={draw,rounded corners,align=center,inner sep=5pt},
 ar/.style={-{Latex[length=2mm]}}]
\node[box] (ref) {Reference and two marginals\\$\EER(\clean),\ \EER(A),\ \EER(B)$};
\node[box,right=10mm of ref] (pred) {Additive prediction\\$\EERhat(A\cap B)$};
\node[box,below=of pred] (joint) {Measured joint error\\$\EER(A\cap B)$};
\draw[ar] (ref) -- (pred);
\draw[ar] (pred) -- node[right] {$\IE,\ \NLI$} (joint);
\node[align=center,below=9mm of ref] {Check population, trials,\\reference terms, and metric scale};
\end{tikzpicture}
\caption{Four terms define an additive comparison. Their numerical availability
permits recomputation; comparable sampling and verified condition provenance
are needed to interpret the contrast.}
\label{fig:protocol}
\end{figure}
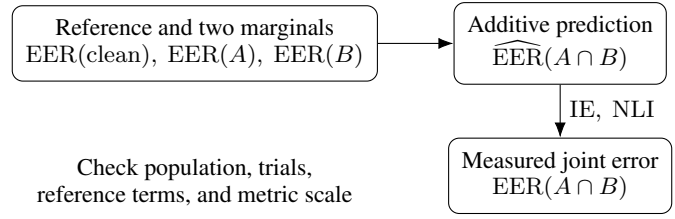

\paragraph{Chance performance and mathematical coupling.}
\label{sec:ceiling}
With fixed score orientation, EER lies in $[0,1]$; $0.5$ is the chance reference,
not a hard upper bound. The supplied routine neither reverses nor clips scores,
and the artefact contains EERs above $0.5$. An additive prediction above chance
therefore does not force a negative interaction. More generally, if an error
metric is bounded above by $U$, then $\IE\leq U-\EERhat$; a prediction above
$U$ necessarily gives a negative contrast. For this EER implementation, $U=1$.

Correlation between $\IE$ and $\EERhat$ is also not an independent test of
saturation: the latter is subtracted in the former. Writing $Y=\EER(A\cap B)$
and $X=\EERhat(A\cap B)$ gives
$\operatorname{Cov}(Y-X,X)=\operatorname{Cov}(Y,X)-\operatorname{Var}(X)$.
A negative correlation can consequently arise without a ceiling mechanism.
We report the additive range and this correlation as descriptive checks,
without using them to certify or exclude an encoder family.

\section{Benchmark Construction}
\label{sec:benchmark}
\paragraph{Corpora and encoder configurations.}
The records name RAVDESS \citep{livingstone2018ravdess}, TORGO \citep{rudzicz2012torgo}, EARS
\citep{richter2024ears}, Common Voice \citep{ardila2020commonvoice}, L2-ARCTIC
\citep{zhao2018l2arctic}, and VoxCeleb1 \citep{nagrani2017voxceleb}.
RIRS\_NOISES is the configured external-noise source
\citep{ko2017reverberation}. The audio front end resamples to 16\,kHz, rejects
clips shorter than 1\,s, and truncates those longer than 10\,s.
The artefact includes speaker counts for some records, but does not provide
complete corpus-release, audio-duration, or sampling manifests. Its trial total sums scoring
operations across conditions and models, not distinct utterances or independent
speaker pairs.

Twelve model labels appear. Seven are assigned to speaker encoders: X-Vector
\citep{snyder2018xvectors}, ECAPA-TDNN and ECAPA-TDNN-Large
\citep{desplanques2020ecapa}, RawNet3 \citep{jung2022rawnet3}, CAM++ \citep{wang2023campp}, TitaNet \citep{koluguri2022titanet}, and ReDimNet
\citep{yakovlev2024redimnet}. The wrappers use
SpeechBrain \citep{ravanelli2021speechbrain}, WeSpeaker
\citep{wang2023wespeaker}, or model-specific loaders, with unrecorded fallback
paths. Thus these labels do not verify seven distinct resolved speaker checkpoints.

Five self-supervised learning (SSL) configurations use WavLM-Base, WavLM-Base+,
WavLM-Large \citep{chen2022wavlm}, UniSpeech-SAT
\citep{chen2022unispeechsat}, and HuBERT-Large \citep{hsu2021hubert}.
Their intended utterance representation is the temporal mean of the final
hidden layer, without a trained speaker head. We retain the recorded family
split for summarisation; it is not a controlled comparison of training paradigms.
Original model-label spellings are retained in tables to permit exact CSV lookup.

\paragraph{Operators and provenance.}
The harness specifies telephony encode--decode operations with \texttt{ffmpeg},
noise at 5, 15, and 25\,dB signal-to-noise ratio (SNR), 20\,ms packet-loss masks,
and time-scale perturbations. The run audit identifies E04 as recorded EARS
styles and E05 as SBC, AAC, aptX and aptX-HD encode--decode conditions.
The DSP/AAC proxy paths in the bundled legacy harness do not describe those
retained runs. Codec round trips do not model a wireless radio channel.
Other transformations also need verification. The environmental and crosstalk
operators draw from the same noise directory, and revert to Gaussian noise when
files are unavailable. Codec failures can return unchanged audio. Synthesis and
conversion wrappers name VoiceCraft \citep{peng2024voicecraft}, E2 TTS
\citep{eskimez2024e2tts}, kNN-VC \citep{baas2023knnvc}, and FreeVC
\citep{li2023freevc}, but also return original audio on failures. The updated export includes augmentation and decode-failure counts for some
records, but these fields do not cover every experiment. Appendix~\ref{app:operators} separates
configured operations from verified execution.

\paragraph{Trials, splits, and leakage.}
The general sampler requests balanced genuine and impostor trials with a cap of
$10^4$, using a module-level random generator initialised with seed 42.
Repeated calls advance that generator; a fixed initial seed does not give the
same trial list to every model or condition. Scoring failures can change class
balance. The general genuine-pair sampler uses different recordings, but this
constraint is not enforced by every specialised sampler. Audio augmentation
randomness and cache state are also not fully captured in the result rows.

No training or validation split is defined for benchmark-specific learning;
the intended protocol freezes pre-trained encoders. This does not establish
that evaluation speakers were unseen in pre-training. Several checkpoints use
VoxCeleb, and E07, E17, and E20 reference VoxCeleb1. Speaker-level overlap has
not been audited. Future model selection on these public conditions would also
require a separate held-out protocol to assess generalisation.

\paragraph{Coverage and availability.}
\label{sec:availability}
The submission CSV contains 4{,}068 scored records from 17 experiments and
12 model labels, totalling 11{,}975{,}026 trial evaluations. It includes ten
single-factor, six two-factor, and one nominal three-factor experiment.
E13 and E14 use only the replacement cohorts, each covering 12 models; the
single-model legacy E13 cohort is excluded. E21 is removed because the updated
cohort covers only one of 13 model labels, leaving the rest as legacy results.
E09, E12, and E19 remain absent. E15 and E18 are restored from the historical
export as descriptive results, with 240 and 120 scored rows respectively. Their
corpus-level counts and aggregate metrics agree with the available run audit;
their cross-corpus contrasts are excluded from the interaction analysis.

The retained roster contains the same 12 models in every experiment, with
complete coverage of each experiment's observed condition grid. ERes2Net is
excluded because it lacks four of the retained experiments. No missing
measurements are imputed. Appendix~\ref{app:inventory}
reports coverage. E05, E10, and E11 contain operator labels in the dataset
field rather than identifiable corpus names; these are shown as unrecorded.

The anonymised supplement contains the selected results, analysis scripts, and
available legacy harness. It permits recomputation of aggregate summaries,
but the harness is not established as the implementation of the updated runs.
No corpus audio is included. Corpus acquisition and reuse remain subject to
original distributor terms; an artefact licence and permanent repository have
not yet been specified.

\paragraph{Intended use.}
\label{sec:usage}
The intended uses are condition-level comparisons, additive contrasts with
verified reference terms, and stress-conditioned subgroup analysis. A low
interaction effect is not itself a robustness score: a system can have low
accuracy under every condition and still have a small contrast. Repeated tuning
on public cells can also overfit the benchmark. Absolute error, balanced
coverage, uncertainty, and held-out conditions are needed alongside interaction
summaries.

\section{Results}
\label{sec:results}
\begin{figure}[t]
\centering
\begin{minipage}[t]{0.48\linewidth}\centering
\includegraphics[width=\linewidth]{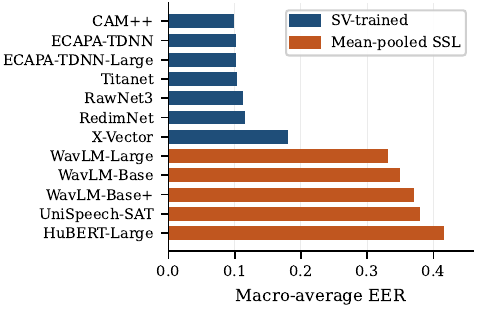}
\end{minipage}\hfill
\begin{minipage}[t]{0.48\linewidth}\centering
\includegraphics[width=\linewidth]{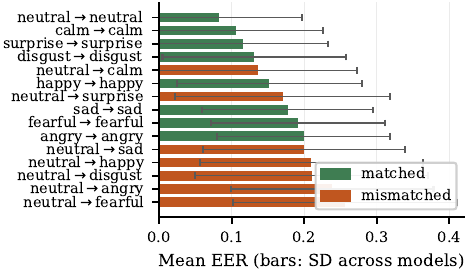}
\end{minipage}
\caption{\textbf{Left:} macro EER by recorded model label; all 12 encoders cover
the same 17 experiments. \textbf{Right:} E01 on RAVDESS, averaging available encoders for each
condition. Bars show between-encoder SD, not uncertainty in a mean. Both panels
use EER fractions. Family assignments and condition labels follow the artefact.}
\label{fig:models}
\end{figure}

\paragraph{Encoder summaries.}
\label{sec:models}
Macro-averaging first averages condition values within experiments and then
weights experiments equally. CAM++ has macro EER 0.098, ECAPA-TDNN
0.101, and ECAPA-TDNN-Large 0.102; nominal speaker-trained labels span 0.098--0.181.
Mean-pooled SSL labels span 0.331--0.415 (Figure~\ref{fig:models};
Appendix~\ref{app:encoders}). All 12 labels cover the same 17 experiments
and each observed condition grid is complete. The macro ordering remains
descriptive: common coverage alone does not establish matched trials or
comparable sampling across encoders. Checkpoint
hashes are present for a subset of the results; sampling and provenance still
limit comparisons.

\paragraph{Emotion and enrolment--test mismatch.}
\label{sec:tier1}
On RAVDESS, the neutral/neutral mean is 0.083 EER and the
neutral$\rightarrow$fearful mean is 0.256, a ratio of $3.11\times$.
Matched fearful speech has mean EER 0.191 ($2.32\times$ the reference).
Each neutral-to-emotion mismatch has a higher mean than the corresponding
same-emotion condition (Appendix~\ref{app:e01}). This pattern is consistent
with a contribution from emotional mismatch, but does not isolate a causal
mechanism or rank emotion against stressors on other corpora. All 15 conditions
cover the same 12 model labels; matched and mismatched trials have different
budgets. Between-encoder SDs are 0.11--0.16, not confidence intervals for the
condition differences. The updated E01 export contains RAVDESS only.

\paragraph{Same-corpus additive contrasts.}
\label{sec:pairwise}
For E13, E14, and E17 we reconstruct Equation~\eqref{eq:ie} from the CSV, retaining
settings with all four finite terms for the same model label. This verifies
the arithmetic, not the sampling or condition provenance
(Section~\ref{sec:identifiability}). Table~\ref{tab:pairwise} retains all such
settings without a post-hoc family-level validity filter.

\begin{table}[t]
\centering
\caption{Descriptive additive contrasts from available same-corpus terms.
Means and SDs are across model--setting combinations; $n$ counts those
combinations, not independent replicates. EER uses fractional units;
``$\IE>0$'' is a percentage. SV denotes nominal speaker-trained labels and SSL
denotes frozen mean-pooled configurations. E13 compares an age stratum with an
overlapping all-age reference; E17's temporal provenance is unresolved.
For E14 SV, $\NLI$ averages 198 positive-denominator cells; $n=441$
includes all contrasts, including 243 with nonpositive predictions.}
\label{tab:pairwise}
\footnotesize\setlength{\tabcolsep}{3pt}
\begin{tabular}{@{}lrrrrrrrrrr@{}}
\toprule
& \multicolumn{5}{c}{\textbf{SV labels}} & \multicolumn{5}{c}{\textbf{SSL labels}} \\
\cmidrule(lr){2-6}\cmidrule(lr){7-11}
Experiment & $\IE$ & SD & $\NLI$ & $\IE>0$ & $n$ &
$\IE$ & SD & $\NLI$ & $\IE>0$ & $n$ \\
\midrule
\input{generated/tab_pairwise.tex}
\end{tabular}
\end{table}

The nominal speaker-trained group has mean contrasts $+0.0026$, $+0.0088$,
and $+0.0024$ for E13, E14, and E17; 52\%, 34\%, and 62\% of combinations
have positive signs. Mean NLIs are 1.089, 1.549, and 1.137. These average
per-cell ratios over positive predictions only; in E14, 243 of the 441 SV
predictions are nonpositive and have no NLI, while their contrasts remain in
the IE summary. Small denominators make NLI particularly unstable in this
experiment. Signed mean departures are small, but SDs of 0.009--0.030 exceed
them. Individual E14 SV contrasts range from $-0.059$ to $+0.139$.
Cancellation and zero contrasts can therefore obscure substantial setting-level
deviations. The table does not establish equivalence to additivity or justify
replacing joint-condition tests with marginal measurements.
Appendices~\ref{app:perencoder} and~\ref{app:settings} report this variation.

\begin{figure}[t]
\centering
\begin{minipage}[t]{0.48\linewidth}\centering
\includegraphics[width=\linewidth]{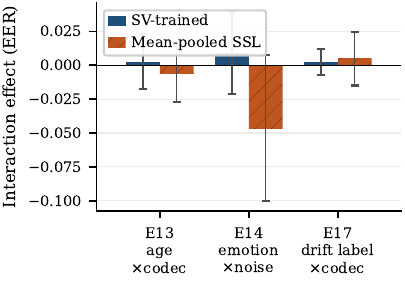}
\end{minipage}\hfill
\begin{minipage}[t]{0.48\linewidth}\centering
\includegraphics[width=\linewidth]{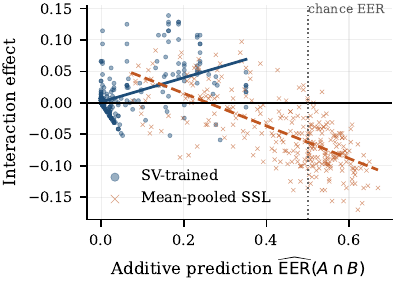}
\end{minipage}
\caption{\textbf{Left:} mean additive contrast by experiment and recorded
family; bars show SD across model--setting combinations. \textbf{Right:} E14
contrasts versus additive predictions, with descriptive least-squares fits.
The dotted line marks chance EER, not an upper bound. The shared prediction
term contributes to the negative correlation.}
\label{fig:ie}
\end{figure}

\paragraph{Near-chance predictions.}
\label{sec:results-ceiling}
For mean-pooled SSL labels in E14, the mean contrast is $-0.0465$, and 22\%
of combinations have positive contrasts. The additive prediction averages
0.438 and reaches 0.670; its correlation with $\IE$ is $r=-0.706$
(Table~\ref{tab:ceiling}, Figure~\ref{fig:ie}). For nominal speaker-trained
labels, the mean prediction is 0.039 and $r=+0.504$. These differences describe
the supplied configurations, without identifying a saturation mechanism:
EER can exceed chance and the correlation shares a term with the contrast.
E13's SSL predictions remain below 0.5, so a chance-ceiling criterion would
not exclude that family either.

\begin{table}[t]
\centering
\caption{Additive-prediction summaries. $\bar{\EERhat}$ and
$\max\EERhat$ are the mean and maximum prediction; $r$ is Pearson correlation
between the prediction and $\IE$. These are descriptive diagnostics, not a test
that distinguishes saturation from mathematical coupling.}
\label{tab:ceiling}
\small
\begin{tabular}{@{}lrrrrrr@{}}
\toprule
& \multicolumn{3}{c}{\textbf{SV labels}} & \multicolumn{3}{c}{\textbf{SSL labels}} \\
\cmidrule(lr){2-4}\cmidrule(lr){5-7}
Exp. & $\bar{\EERhat}$ & $\max\EERhat$ & $r$ &
$\bar{\EERhat}$ & $\max\EERhat$ & $r$ \\
\midrule
\input{generated/tab_ceiling.tex}
\end{tabular}
\end{table}

Across the five SSL labels in E14, mean predictions range from 0.397 to 0.457,
observed means from 0.350 to 0.414, and mean contrasts from $-0.073$ to
$-0.028$. This within-family variation is descriptive, rather than a control
that separates competing explanations for negative contrasts.

\begin{figure}[t]
\centering
\includegraphics[width=0.66\linewidth]{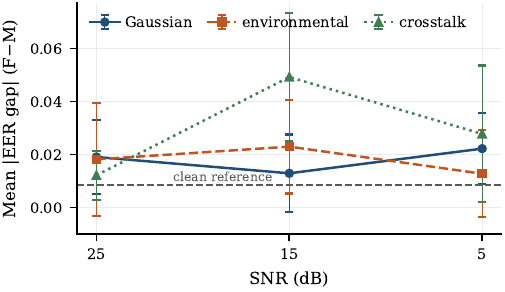}
\caption{E16 mean absolute gender EER gap by labelled SNR. Bars show SD across
12 model labels. Noise names follow the CSV; source-class provenance is
unverified. The dashed line is the clean mean. Independently sampled trial
lists and unequal clean/noisy budgets limit causal interpretation.}
\label{fig:gendergap}
\end{figure}

\paragraph{Demographic gaps under stress.}
\label{sec:fairness}
E16 reports gender-stratified Common Voice conditions. We average the absolute
per-model gap $|\EER_F-\EER_M|$ over 12 model labels. The mean gap is 0.0086
in the clean condition. At 25, 15, and 5\,dB, Gaussian-labelled noise gives
0.0191, 0.0129, and 0.0222; environmental-labelled noise gives 0.0182, 0.0230,
and 0.0128; and crosstalk gives 0.0122, 0.0493, and 0.0278.
All nine noisy means exceed the clean mean, but none of the three sequences
increases monotonically as SNR falls (Figure~\ref{fig:gendergap}). This does
not imply an effect for every encoder or identify one population as consistently
worse served. Appendix~\ref{app:gendergap} lists the subgroup summaries.

The signed clean gap ranges in magnitude from 0.0004 to 0.0352 across models.
Dividing noisy signed gaps by these small values produces ratios from $-157.6$
to $+193.7$ (Appendix~\ref{app:baf}). Absolute gaps avoid that denominator,
but still need sampling uncertainty: E16 uses 5{,}000 clean versus 555 noisy
trials per subgroup, and absolute differences can increase with estimator
variability. Three clean signed gaps are positive and nine negative, so there
is no uniform direction across encoders.

\section{Discussion}
\label{sec:discussion}
An additive baseline makes the comparison of compound and single conditions
explicit. The small signed means in the artefact motivate testing its predictive
accuracy, but do not establish it. A useful validation would predefine an
acceptable error, measure prediction errors on held-out factor settings, and
retain their per-model distribution. A benchmark should report where the
prediction fails, not only whether positive and negative deviations cancel.

Metric interpretation is equally consequential. EER evaluates a condition's
own equal-error operating point; it does not directly measure risk at a fixed
production threshold. Likewise, demographic EER gaps compare separately
optimised operating points. Deployment-oriented analysis needs subgroup
false-accept and false-reject rates at a common threshold selected on separate
development data. These additions would test the practical meaning of the
observed contrasts without treating benchmark averages as deployment guarantees.

\section{Limitations and Threats to Validity}
\label{sec:limitations}
\textbf{Result provenance.} The selected CSV includes run IDs for 2{,}508
records, revision identifiers for 2{,}130, checkpoint hashes for 1{,}878, and
genuine-speaker counts for 924. Coverage is incomplete; trial lists, per-trial
scores, hardware, and full software environments are not supplied. The legacy
harness can substitute checkpoints under other labels, whereas the updated
E13/E14 cohorts carry explicit hashes. Aggregate agreement alone cannot
authenticate an encoder (Appendix~\ref{app:agreement}).

\textbf{Code--artefact discrepancies.} The bundled legacy harness differs from
producing revisions. The run audit found positive-gap execution, two-class
E11 scoring, recorded E04 styles, real E05 codecs, and measured E20 reference
terms. These findings supersede failures inferred solely from old code paths.
Exact scored-pair manifests and full checkpoint binding are still unavailable;
publication-year and age proxies, and the unrestricted nominal zero-gap
reference, limit temporal interpretation (Appendix~\ref{app:limitations}).

\textbf{Uncertainty and comparability.} Each retained record has one aggregate
measurement and no linked score list. Repeated speakers and shared reference
terms make the contrasts dependent. Between-model SDs do not quantify sampling
uncertainty. Trial budgets vary, and corpus membership alone
does not establish matched population composition.

\textbf{Scope.} Per-trial noise, codec, synthesis, and conversion execution
is not completely documented. E15/E18 mix source populations across corpora;
E20 uses temporal and age proxies. Pre-training overlap, corpus versions, full subgroup sizes,
and metadata provenance remain unaudited. These limits prevent broad claims
about real-world compound robustness or demographic fairness.

\section{Conclusion}
\label{sec:conclusion}
InterBias-SV organises compound-condition SV evaluation around measured
reference terms and an explicit additive contrast. The supplied artefact permits
recomputable summaries for three same-corpus experiment families, with small
positive signed means for nominal speaker-trained labels and substantial
setting-level variation. Those averages remain descriptive until trial matching
and experimental provenance are established. Chance EER is not a hard ceiling,
and a near-zero demographic reference makes gap ratios unstable. The framework
and audit identify what must be measured and recorded before compound-condition
results can support stronger conclusions.

\FloatBarrier
\clearpage
\subsection*{Ethics Statement}
The benchmark uses existing speech corpora and introduces no new participant
recruitment or annotation. The supplementary package contains aggregate results
and code, without corpus audio. Public availability does not itself establish
consent for every downstream use: corpus versions, licence terms, and permitted
uses must be checked with the original distributors. No claim of institutional
ethics exemption follows from the absence of new data collection.

Demographic categories are coarse, subgroup metadata are incomplete, and the
retained gender and age comparisons do not establish fairness for specific
populations. Speaker counts in some updated rows do not resolve label provenance
or sampling uncertainty. The results should not be used to certify
fairness or to justify deployment of biometric surveillance or access-control
systems. Speech and linked demographic metadata can reveal identity even when
speaker IDs are pseudonymous; any later release of manifests or audio-derived
information needs a separate privacy and licence review. Spoofing and conversion
experiments concern impersonation risks, but their execution must be verified
before drawing vulnerability conclusions. No generated attack audio is included.

\subsection*{Reproducibility Statement}
Section~\ref{sec:protocol} defines scoring and contrasts;
Section~\ref{sec:benchmark} and Appendix~\ref{app:operators} describe the
available implementation. The supplementary CSV and two analysis scripts
reconstruct aggregate tables and plots, with detailed listings in
Appendices~\ref{app:interaction} and~\ref{app:conditions}. The protocol diagram
is defined in the LaTeX source. This is summary-level reproducibility:
Section~\ref{sec:limitations} and Appendix~\ref{app:limitations} identify missing
run provenance and code--artefact discrepancies that prevent full experimental
reproduction.

\subsection*{AI Use Statement}
Generative AI assisted implementation and analysis scripting, results-file
formatting, interpretation of numerical summaries, literature retrieval,
reference formatting, and manuscript drafting and editing. It also provided
methodological criticism, suggested follow-up experiments, and helped revise
mathematical statements about EER and interaction contrasts during the
submission-readiness review. The supplied retained observations were not altered; this editorial
review did not generate experimental observations or run new benchmark
experiments. Checks in this revision covered aggregate recomputation, source-code
consistency, selected primary references, and document compilation. These checks
do not establish the provenance of the underlying experiments. The authors
remain responsible for verifying the scientific content and for ensuring that
the submission-form disclosure accurately reflects the full research workflow.

\bibliography{references}
\bibliographystyle{iclr2027_conference}
\clearpage
\appendix
\input{appendix/appendix.tex}

\end{document}

%% file: generated/tab_pairwise.tex
E13 age $\times$ codec & +0.0026 & 0.0204 & 1.089 & 52 & 84 & -0.0063 & 0.0209 & 0.973 & 35 & 60 \\
E14 emotion $\times$ noise & +0.0088 & 0.0301 & 1.549 & 34 & 441 & -0.0465 & 0.0540 & 0.925 & 22 & 315 \\
E17 drift $\times$ codec & +0.0024 & 0.0094 & 1.137 & 62 & 84 & +0.0049 & 0.0198 & 1.014 & 57 & 60 \\
\bottomrule%

%% file: generated/tab_ceiling.tex
E13 & 0.053 & 0.125 & +0.070 & 0.254 & 0.399 & +0.012 \\
E14 & 0.039 & 0.351 & +0.504 & 0.438 & 0.670 & -0.706 \\
E17 & 0.048 & 0.208 & -0.220 & 0.392 & 0.467 & -0.293 \\
\bottomrule%

%% file: appendix/appendix.tex
\FloatBarrier
\section{Benchmark Documentation}
\label{app:doc}

\FloatBarrier
\subsection{Motivation, composition and intended use}
\label{app:operators}

\paragraph{Purpose.} InterBias-SV specifies additive comparisons between joint
conditions and their measured reference terms. The results below reproduce the
supplied aggregate artefact. Numerical availability is distinct from validated
condition provenance and matched sampling (Section~\ref{sec:identifiability}).

\paragraph{Instances.} A condition combines a source population, transformation, and trial-construction
rule. The selected file contains 4{,}068 records with EER, minDCF, AUC, and trial
counts across 17 experiments, totalling 11{,}975{,}026 trial evaluations. This
sum includes reuse across conditions and models. All included records have
finite scores; absent model--condition cells are not imputed.

\paragraph{Degradation operators.} Table~\ref{tab:operators} documents the
available implementation, not proof that each operation succeeded in the runs
that produced the CSV. Failure counts are supplied for some updated records, but execution provenance
is incomplete.

\begin{table}[htbp]
\centering
\caption{Legacy harness operators (not the producing revisions for all retained rows). Proxy operators are marked $p$.
Execution and checkpoint provenance are not verified by the result rows.}
\label{tab:operators}
\footnotesize
\setlength{\tabcolsep}{4pt}
\begin{tabular}{@{}p{1.9cm}p{3.1cm}p{7.5cm}@{}}
\toprule
\textbf{Operator} & \textbf{Settings} & \textbf{Implementation} \\
\midrule
Telephony codec & G.711, GSM 06.10, AMR-NB, Opus &
  \texttt{ffmpeg} encode--decode round trip; G.711 $\mu$-law, GSM and AMR-NB at
  8\,kHz, Opus at 16\,kHz / 12\,kbit/s; failures return original audio \\
Additive noise & Gaussian, environmental, crosstalk; 5/15/25\,dB SNR &
  environmental and crosstalk labels share a RIRS\_NOISES directory
  \citep{ko2017reverberation}; missing files trigger Gaussian fallback \\
Packet loss & 0.5\%, 2\%, 5\%, 10\% &
  20\,ms frames zeroed independently; no packet-loss concealment or burst-loss model \\
Speaking rate & $0.75\times$, $1.0\times$, $1.25\times$ & time-scale perturbation \\
Wireless codec$^{p}$ & 32--990\,kbit/s &
  every condition encoded with AAC at a varying bitrate under labels nominally
  denoting SBC, aptX and LDAC; measures perceptual-codec bitrate sensitivity,
  not codec identity \\
Voice mode$^{p}$ & whisper, shout &
  signal-level DSP proxy, not recorded phonation-mode speech; EARS contains
  genuine whispered and shouted recordings, which this operator does not use \\
Synthesis & VoiceCraft \citep{peng2024voicecraft}, E2-TTS
  \citep{eskimez2024e2tts} & named wrappers; failures can return original audio \\
Voice conversion & kNN-VC \citep{baas2023knnvc}, FreeVC
  \citep{li2023freevc} & named wrappers; failures can return original audio \\
\midrule
Audio front end & 16\,kHz; 1--10\,s & clips $<1$\,s discarded, clips $>10$\,s truncated \\
Trials & $\leq 10^{4}$ per condition & balanced sampling requested; module-level seed 42; condition-specific
  resampling and exclusions; specialised samplers need separate checks \\
Scoring & cosine on $L_2$-normalised embeddings &
  $\mindcf$ at $P_{\text{target}}=0.01$, $C_{\text{miss}}=C_{\text{fa}}=1$; no
  score normalisation, calibration or adaptation \\
\bottomrule
\end{tabular}
\end{table}

\paragraph{Splits.} There is no benchmark-specific training split or fine-tuning.
Pre-trained encoders may nevertheless have seen evaluation speakers. Corpus
versions, sampling manifests, and pre-training overlap need to be established
before any condition can be described as held out.

\paragraph{Intended use.} The artefact supports descriptive condition summaries
and recomputation of same-corpus contrasts. Controlled interaction, demographic,
spoofing, and temporal-drift claims require the provenance and experimental
checks described in Appendix~\ref{app:limitations}.

\FloatBarrier
\subsection{Full experiment inventory with aggregate scores}
\label{app:inventory}

Table~\ref{tab:inventory} is the per-experiment inventory: corpus, encoder and
condition counts, scored records, trial counts, and the descriptive use of each experiment under
Section~\ref{sec:identifiability}. Table~\ref{tab:coverage-full} adds the
aggregate EER, minDCF and AUC per experiment. Those aggregates pool
heterogeneous conditions and encoders: they describe the difficulty of a
condition set and are not a controlled ranking of stressors.

\begin{table}[htbp]
\centering
\caption{Experiment inventory. ``Scored'' counts model--condition records with a
computed EER. ``Use'' is what the experiment is admitted for:
$\dagger$~cross-corpus populations: descriptive only; $\ddagger$~temporal/age
proxy interpretation; $\S$~crosses a population attribute
with a stressor, analysed as a subgroup gap; $p$~legacy proxy implementation only. E05, E10, and E11 use operator labels rather than corpus identifiers in their
dataset fields. All 12 retained encoders have complete coverage of each observed condition grid.}
\label{tab:inventory}
\scriptsize
\setlength{\tabcolsep}{3pt}
\begin{tabular}{@{}llp{2.15cm}p{1.8cm}rrrrp{1.8cm}@{}}
\toprule
\textbf{T} & \textbf{ID} & \textbf{Factor(s)} & \textbf{Corpus} & \textbf{Enc.} &
\textbf{Cond.} & \textbf{Scored} & \textbf{Trials} & \textbf{Use} \\
\midrule
\input{generated/tab_inventory.tex}
\end{tabular}
\end{table}

\begin{table}[htbp]
\centering
\caption{Experiment inventory with aggregate scores. EER is the unweighted mean
over scored records with its standard deviation across those records.
$\dagger$~cross-corpus populations, reported descriptively;
$\ddagger$~temporal/age proxy interpretation.}
\label{tab:coverage-full}
\footnotesize
\setlength{\tabcolsep}{3pt}
\begin{tabular}{@{}llp{2.1cm}rrrrrrr@{}}
\toprule
\textbf{Tier} & \textbf{ID} & \textbf{Factor(s)} & \textbf{Enc.} &
\textbf{Cond.} & \textbf{Scored} & \textbf{Trials} & \textbf{EER} &
\textbf{minDCF} & \textbf{AUC} \\
\midrule
\input{generated/tab_coverage.tex}
\end{tabular}
\end{table}

\paragraph{Historical descriptive experiments (E15 and E18).}
E15 contributes 240 scored rows (12 encoders, 20 conditions): TORGO dysarthric
clean/codec results and EARS healthy-reference results. E18 contributes 120 rows
(12 encoders, 10 conditions): L2-ARCTIC and Common Voice English clean/codec
results. All condition grids are complete. Compare clean and codec conditions
within each corpus; differences between corpora cannot isolate pathology or
accent. E18's literal \texttt{native} condition labels are retained for lookup,
but English locale does not establish native-speaker status. These historical
rows enter descriptive summaries only, not the E13/E14/E17 interaction table.

\FloatBarrier
\section{Encoder Results}
\label{app:encoders}

Table~\ref{tab:models} gives macro-average results: conditions are averaged
within each experiment, then experiments are weighted equally. The last column
is the number of experiments contributing: seventeen for every retained encoder.
Condition coverage is complete, but trial matching and provenance limitations
keep the ordering descriptive.

\begin{table}[htbp]
\centering
\caption{Macro-average encoder results. ``Exp.'' is the number of experiments
contributing to that encoder's macro average.}
\label{tab:models}
\small
\begin{tabular}{@{}llrrrr@{}}
\toprule
\textbf{Encoder} & \textbf{Family} & \textbf{EER} & \textbf{minDCF} &
\textbf{AUC} & \textbf{Exp.} \\
\midrule
\input{generated/tab_models.tex}
\end{tabular}
\end{table}

\FloatBarrier
\section{Single-Factor Results: E01 Emotional Speech}
\label{app:e01}

\begin{table}[htbp]
\centering
\caption{E01, RAVDESS. Mean EER over available encoders, with the standard
deviation across encoders; the last column is the ratio to the neutral/neutral
reference. ``Matched'' conditions use the same emotion for enrolment and test.
The SD is between-encoder variability, not a confidence interval. All conditions cover the same 12 model labels. Ratios divide the condition
mean by the neutral-reference mean; they do not quantify sampling uncertainty.}
\label{tab:e01}
\small
\setlength{\tabcolsep}{5pt}
\begin{tabular}{@{}llrrrrr@{}}
\toprule
\textbf{Condition} & \textbf{Type} & \textbf{EER} & \textbf{SD} &
\textbf{minDCF} & \textbf{AUC} & \textbf{Ratio} \\
\midrule
\input{generated/tab_e01.tex}
\end{tabular}
\end{table}

\FloatBarrier
\section{Demographic Results}
\label{app:demographic}

\FloatBarrier
\subsection{Absolute gender gap by condition}
\label{app:gendergap}

\begin{table}[htbp]
\centering
\caption{E16 gender gap on Common Voice. F and M are the recorded subgroup labels;
columns give mean EERs over 12 model labels; $|$gap$|$ is the mean absolute per-encoder difference, and
max$|$gap$|$ its largest value over encoders.}
\label{tab:gendergap}
\small
\setlength{\tabcolsep}{5pt}
\begin{tabular}{@{}lrrrrr@{}}
\toprule
\textbf{Condition} & \textbf{EER (F)} & \textbf{EER (M)} &
\textbf{$|$gap$|$} & \textbf{max$|$gap$|$} & \textbf{Enc.} \\
\midrule
\input{generated/tab_gendergap.tex}
\end{tabular}
\end{table}

\FloatBarrier
\subsection{Instability of ratio-form bias amplification}
\label{app:baf}

\begin{table}[htbp]
\centering
\caption{Instability of ratio-form bias amplification on E16. The clean gender
gap is the denominator of the ratio; ``mean $|$gap$|$'' is the mean absolute
gap over the nine noise conditions. The ratio range is reported to show its
sensitivity to a near-zero denominator, not as a reliable amplification estimate.}
\label{tab:baf}
\small
\begin{tabular}{@{}lrrr@{}}
\toprule
\textbf{Encoder} & \textbf{Clean gap} & \textbf{Mean $|$gap$|$} &
\textbf{Ratio range} \\
\midrule
\input{generated/tab_baf.tex}
\end{tabular}
\end{table}

\FloatBarrier
\section{Interaction Analysis in Detail}
\label{app:interaction}

\FloatBarrier
\subsection{Per-encoder interaction effects}
\label{app:perencoder}

Table~\ref{tab:perenc} decomposes the contrasts by model label. In E14,
all five SSL means are negative, ranging from $-0.073$ to $-0.028$ EER.
Six of seven speaker-trained means are positive; RawNet3 is negative
($-0.003$). The table includes all finite four-term contrasts, while NLI means
use only positive additive predictions. No family-level chance screen is applied;
sampling and provenance limitations remain.

{\footnotesize\setlength{\tabcolsep}{4pt}
\begin{longtable}{@{}llrrrrrrr@{}}
\caption{Interaction effect by encoder and experiment, over the settings for
which all four factorial terms are available. $n$ is the number of
settings.}\label{tab:perenc}\\
\toprule Exp. & Encoder & Fam. & $\overline{\IE}$ & SD & min & max &
$\overline{\NLI}$ & $n$ \\
\midrule\endfirsthead
\multicolumn{9}{c}{\tablename~\thetable{} -- continued}\\
\toprule Exp. & Encoder & Fam. & $\overline{\IE}$ & SD & min & max &
$\overline{\NLI}$ & $n$ \\
\midrule\endhead
\bottomrule\endfoot
\input{generated/appendix/s2_per_encoder_ie.tex}
\end{longtable}}

\FloatBarrier
\subsection{Setting-level interaction listing}
\label{app:settings}

Table~\ref{tab:settings} gives the pooled observed joint EER and additive prediction, followed by separate
family-mean contrasts for every setting with four available terms. The dispersion visible here is what the
pooled means of Section~\ref{sec:pairwise} average over.

{\footnotesize\setlength{\tabcolsep}{4pt}
\begin{longtable}{@{}lp{0.30\textwidth}rrrr@{}}
\caption{Per-setting interaction summary. ``Obs.'' and ``Add.'' pool all
available encoder labels; the final two columns give family-specific mean
contrasts.}\label{tab:settings}\\
\toprule Exp. & Setting & Obs. (all) & Add. (all) & $\IE$ (SV) & $\IE$ (SSL) \\
\midrule\endfirsthead
\multicolumn{6}{c}{\tablename~\thetable{} -- continued}\\
\toprule Exp. & Setting & Obs. (all) & Add. (all) & $\IE$ (SV) & $\IE$ (SSL) \\
\midrule\endhead
\bottomrule\endfoot
\input{generated/appendix/s3_settings.tex}
\end{longtable}}

\FloatBarrier
\section{Encoder-Agreement Audit}
\label{app:agreement}

Several model wrappers can replace a requested encoder with a SpeechBrain
ECAPA-TDNN checkpoint, without recording the resolved identity in the CSV.
Table~\ref{tab:agreement} compares aggregate rows. No pair agrees exactly on
more than 20.9\% of the compared conditions, but this cannot exclude fallback:
models can receive different sampled trials and augmentation realisations.
Agreement rates include exactly zero errors in the updated matched conditions
and cannot authenticate a checkpoint. The provided hashes offer additional
provenance for the subset of records that contains them.

\begin{table}[htbp]
\centering
\caption{The twelve most similar encoder pairs of 66, by exact agreement over
co-observed conditions. ``Cond.'' is the number of co-observed conditions,
``Agree'' the percentage on which the two EERs are identical, $\rho$ the
correlation of their EERs, and MAD the mean absolute EER difference.}
\label{tab:agreement}
\small
\begin{tabular}{@{}llrrrr@{}}
\toprule
\textbf{Encoder A} & \textbf{Encoder B} & \textbf{Cond.} & \textbf{Agree (\%)} &
$\rho$ & \textbf{MAD} \\
\midrule
\input{generated/appendix/s4_agreement.tex}
\end{tabular}
\end{table}

\FloatBarrier
\section{Extended Limitations}
\label{app:limitations}

\paragraph{Uncertainty.} A single run does not in principle prevent uncertainty
estimation from retained scores. Here, only aggregate measurements are supplied.
Recover per-trial scores and identities, then use a sampling scheme that respects
repeated speakers and shared references across the four terms. Independent
speaker/trial and augmentation seeds would separately test construction
sensitivity. Treating all model--setting combinations as independent replicates
would understate dependence. An equivalence claim additionally needs a
predefined practical margin; a small signed mean is insufficient.

\paragraph{Checkpoint and run provenance.} Preserve the resolved model identifier,
checkpoint hash, software environment, and successful operator configuration
for each run. Disable silent substitution in future measurements. The updated summary includes run identifiers, revision fields, and checkpoint
hashes for subsets of records, but no score manifests linking all observations
to the bundled legacy implementation. The agreement table cannot fill this gap.

\paragraph{Temporal labels and producing revisions.} The run audit found
per-video years, positive-gap execution and measured E20 reference terms.
The old bundled loader is not the producing revision. Publication dates and
age estimates remain proxies; the nominal zero-gap reference is unrestricted,
not a verified same-session or same-year baseline. Complete pair manifests
are needed to verify sampling and overlap.

\paragraph{Noise and attack operators.} The audit found two-class E11 scoring
and stored synthesis/conversion outputs. This resolves contradictions inferred
from the legacy one-class branch, but does not supply exact scored-pair lists
or establish real-time latency. Noise-source corrections and failure counts
cover some runs; complete operator and checkpoint binding remains unavailable.

\paragraph{Demographic and emotion labels.} E13 compares an age stratum with an
overlapping all-age reference, not two disjoint age populations. Common Voice
age bins are mapped to representative integers rather than measured ages.
Report speaker counts, missing labels, label sources, and mapping rules for all
strata. The retained E01 and replacement E14 rows use RAVDESS. The paired and matched
cohort labels do not replace trial manifests needed to verify population and
content matching.

\paragraph{Contamination and coverage.} Check pre-training/evaluation speaker
overlap for each resolved checkpoint, especially on VoxCeleb1. All twelve retained labels cover all seventeen experiments and every observed
condition. Comparable speaker composition and successful trials still need
verification; complete coverage does not establish matched sampling.

\paragraph{Excluded experiments and cohorts.} E21 is omitted: only one model
has an updated cohort, with the other twelve represented by legacy rows.
The single-model legacy E13 cohort is omitted in favour of the 12-model paired
replacement. E09, E12, and E19 have no valid replacement results.
ERes2Net is excluded because it lacks four retained experiments.
E15 and E18 are restored only as historical descriptive results. E20 remains descriptive and is not used as a three-way
interaction estimate.

\paragraph{Scale and release metadata.} The trial total counts scoring operations,
not independent data. The release lacks complete corpus versions, speaker-count coverage, audio hours,
run hardware, exact library versions, and execution costs. Some records include
checkpoint hashes and speaker counts. E05, E10, and E11 lack usable corpus identifiers. An artefact licence,
third-party attribution, acquisition instructions, and stable protocol manifests
are needed for reuse.

\FloatBarrier
\section{Condition-Level Results for Every Experiment}
\label{app:conditions}

The tables below list every scored condition in the selected 17 experiments,
sorted by mean EER. ``SD'' is the standard deviation across contributing model
rows, not a confidence interval. There are no unscored rows in the selected
file. Every observed condition has all twelve retained encoders; the encoder
count in each row makes this coverage explicit. CV denotes Common Voice, and
L2-ARCTIC/CV denotes a condition involving both corpora.

The E13 and E14 identifiers correspond to the paired-age and matched-RAVDESS
replacement exports. Their double-underscore condition strings are preserved
for exact lookup. The run audit identifies recorded EARS styles for E04 and actual codec
round trips for E05; the proxy paths in the bundled legacy harness do not
describe those retained observations.

\input{generated/appendix/s1_conditions.tex}

%% file: generated/tab_inventory.tex
1 & E01 & Emotional speech & RAVDESS & 12 & 15 & 180 & 330,000 & marginal \\
1 & E02 & Dysarthric speech & TORGO & 12 & 4 & 48 & 153,706 & marginal \\
1 & E03 & Speaking rate & CommonVoice, EARS & 12 & 18 & 216 & 2,159,910 & marginal \\
1 & E04 & Recorded voice style & EARS & 12 & 9 & 108 & 117,036 & marginal \\
1 & E05 & Codec round trips & \emph{not recorded} & 12 & 15 & 180 & 1,799,550 & marginal \\
1 & E06 & Packet loss & CommonVoice, EARS & 12 & 18 & 216 & 2,159,901 & marginal \\
1 & E07 & Temporal drift & VoxCeleb1 & 12 & 4 & 48 & 480,000 & marginal \\
1 & E08 & L2 accent & CommonVoice, L2-ARCTIC & 12 & 18 & 216 & 378,840 & marginal \\
1 & E10 & TTS spoofing & \emph{not recorded} & 12 & 7 & 84 & 815,448 & marginal \\
1 & E11 & Voice conversion & \emph{not recorded} & 12 & 3 & 36 & 180,000 & marginal \\
2 & E13 & Age $\times$ codec & CommonVoice & 12 & 20 & 240 & 1,199,940 & contrast (audit) \\
2 & E14 & Emotion $\times$ noise & RAVDESS & 12 & 80 & 960 & 276,480 & contrast (audit) \\
2 & E15 & Pathology $\times$ codec$^{\dagger}$ & EARS, TORGO & 12 & 20 & 240 & 468,067 & descriptive$^{\dagger}$ \\
2 & E16 & Gender $\times$ noise & CommonVoice & 12 & 30 & 360 & 419,868 & subgroup gap$^{\S}$ \\
2 & E17 & Drift $\times$ codec & VoxCeleb1 & 12 & 20 & 240 & 479,760 & contrast (audit) \\
2 & E18 & Accent $\times$ codec$^{\dagger}$ & CommonVoice, L2-ARCTIC & 12 & 10 & 120 & 239,648 & descriptive$^{\dagger}$ \\
3 & E20 & Age $\times$ codec $\times$ drift$^{\ddagger}$ & VoxCeleb1 & 12 & 48 & 576 & 316,872 & descriptive$^{\ddagger}$ \\
\bottomrule%

%% file: generated/tab_coverage.tex
1 & E01 & Emotional speech & 12 & 15 & 180 & 330,000 & 0.172 $\pm$ 0.137 & 0.617 & 0.880 \\
1 & E02 & Dysarthric speech & 12 & 4 & 48 & 153,706 & 0.318 $\pm$ 0.109 & 0.964 & 0.737 \\
1 & E03 & Speaking rate & 12 & 18 & 216 & 2,159,910 & 0.351 $\pm$ 0.132 & 0.887 & 0.688 \\
1 & E04 & Recorded voice style & 12 & 9 & 108 & 117,036 & 0.182 $\pm$ 0.183 & 0.531 & 0.850 \\
1 & E05 & Codec round trips & 12 & 15 & 180 & 1,799,550 & 0.252 $\pm$ 0.161 & 0.686 & 0.791 \\
1 & E06 & Packet loss & 12 & 18 & 216 & 2,159,901 & 0.188 $\pm$ 0.162 & 0.582 & 0.853 \\
1 & E07 & Temporal drift & 12 & 4 & 48 & 480,000 & 0.175 $\pm$ 0.183 & 0.524 & 0.854 \\
1 & E08 & L2 accent & 12 & 18 & 216 & 378,840 & 0.136 $\pm$ 0.169 & 0.413 & 0.891 \\
1 & E10 & TTS spoofing & 12 & 7 & 84 & 815,448 & 0.407 $\pm$ 0.147 & 0.876 & 0.620 \\
1 & E11 & Voice conversion & 12 & 3 & 36 & 180,000 & 0.274 $\pm$ 0.128 & 0.737 & 0.774 \\
2 & E13 & Age $\times$ codec & 12 & 20 & 240 & 1,199,940 & 0.130 $\pm$ 0.110 & 0.505 & 0.913 \\
2 & E14 & Emotion $\times$ noise & 12 & 80 & 960 & 276,480 & 0.178 $\pm$ 0.195 & 0.439 & 0.846 \\
2 & E15 & Pathology $\times$ codec$^{\dagger}$ & 12 & 20 & 240 & 468,067 & 0.311 $\pm$ 0.119 & 0.883 & 0.739 \\
2 & E16 & Gender $\times$ noise & 12 & 30 & 360 & 419,868 & 0.182 $\pm$ 0.156 & 0.540 & 0.863 \\
2 & E17 & Drift $\times$ codec & 12 & 20 & 240 & 479,760 & 0.188 $\pm$ 0.177 & 0.560 & 0.847 \\
2 & E18 & Accent $\times$ codec$^{\dagger}$ & 12 & 10 & 120 & 239,648 & 0.145 $\pm$ 0.144 & 0.529 & 0.893 \\
3 & E20 & Age $\times$ codec $\times$ drift$^{\ddagger}$ & 12 & 48 & 576 & 316,872 & 0.186 $\pm$ 0.176 & 0.524 & 0.850 \\
\bottomrule%

%% file: generated/tab_models.tex
CAM++ & SV & 0.098 & 0.347 & 0.933 & 17 \\
ECAPA-TDNN & SV & 0.101 & 0.360 & 0.932 & 17 \\
ECAPA-TDNN-Large & SV & 0.102 & 0.357 & 0.931 & 17 \\
Titanet & SV & 0.104 & 0.341 & 0.928 & 17 \\
RawNet3 & SV & 0.113 & 0.409 & 0.917 & 17 \\
RedimNet & SV & 0.115 & 0.381 & 0.919 & 17 \\
X-Vector & SV & 0.181 & 0.637 & 0.877 & 17 \\
WavLM-Large & SSL & 0.331 & 0.949 & 0.720 & 17 \\
WavLM-Base & SSL & 0.350 & 0.927 & 0.695 & 17 \\
WavLM-Base+ & SSL & 0.371 & 0.961 & 0.673 & 17 \\
UniSpeech-SAT & SSL & 0.380 & 0.967 & 0.661 & 17 \\
HuBERT-Large & SSL & 0.415 & 0.986 & 0.617 & 17 \\
\bottomrule%

%% file: generated/tab_e01.tex
neutral/neutral & matched & 0.083 & 0.114 & 0.424 & 0.951 & 1.00$\times$ \\
calm/calm & matched & 0.106 & 0.119 & 0.378 & 0.936 & 1.28$\times$ \\
surprise/surprise & matched & 0.116 & 0.117 & 0.501 & 0.927 & 1.40$\times$ \\
disgust/disgust & matched & 0.131 & 0.127 & 0.513 & 0.918 & 1.59$\times$ \\
neutral$\rightarrow$calm & mismatched & 0.136 & 0.137 & 0.527 & 0.905 & 1.65$\times$ \\
happy/happy & matched & 0.151 & 0.128 & 0.580 & 0.899 & 1.84$\times$ \\
neutral$\rightarrow$surprise & mismatched & 0.170 & 0.148 & 0.709 & 0.879 & 2.06$\times$ \\
sad/sad & matched & 0.177 & 0.118 & 0.603 & 0.881 & 2.14$\times$ \\
fearful/fearful & matched & 0.191 & 0.120 & 0.589 & 0.869 & 2.32$\times$ \\
angry/angry & matched & 0.199 & 0.119 & 0.671 & 0.870 & 2.41$\times$ \\
neutral$\rightarrow$sad & mismatched & 0.200 & 0.139 & 0.704 & 0.856 & 2.42$\times$ \\
neutral$\rightarrow$happy & mismatched & 0.210 & 0.154 & 0.768 & 0.842 & 2.54$\times$ \\
neutral$\rightarrow$disgust & mismatched & 0.210 & 0.160 & 0.711 & 0.843 & 2.55$\times$ \\
neutral$\rightarrow$angry & mismatched & 0.238 & 0.140 & 0.777 & 0.819 & 2.89$\times$ \\
neutral$\rightarrow$fearful & mismatched & 0.256 & 0.155 & 0.798 & 0.799 & 3.11$\times$ \\
\bottomrule%

%% file: generated/tab_gendergap.tex
clean (reference) & 0.110 & 0.117 & 0.0086 & 0.0352 & 12 \\
Gaussian, 25~dB & 0.139 & 0.141 & 0.0191 & 0.0595 & 12 \\
Gaussian, 15~dB & 0.139 & 0.146 & 0.0129 & 0.0432 & 12 \\
Gaussian, 5~dB & 0.159 & 0.180 & 0.0222 & 0.0450 & 12 \\
environmental, 25~dB & 0.131 & 0.147 & 0.0182 & 0.0811 & 12 \\
environmental, 15~dB & 0.171 & 0.187 & 0.0230 & 0.0523 & 12 \\
environmental, 5~dB & 0.201 & 0.200 & 0.0128 & 0.0631 & 12 \\
crosstalk, 25~dB & 0.167 & 0.162 & 0.0122 & 0.0324 & 12 \\
crosstalk, 15~dB & 0.270 & 0.236 & 0.0493 & 0.0829 & 12 \\
crosstalk, 5~dB & 0.347 & 0.344 & 0.0278 & 0.0883 & 12 \\
\bottomrule%

%% file: generated/tab_baf.tex
CAM++ & -0.0022 & 0.0208 & -37.7 to +8.2 \\
ECAPA-TDNN & -0.0076 & 0.0158 & -5.2 to +3.8 \\
ECAPA-TDNN-Large & -0.0106 & 0.0194 & -6.3 to +1.9 \\
HuBERT-Large & +0.0036 & 0.0198 & -17.5 to +3.0 \\
RawNet3 & -0.0048 & 0.0164 & -6.4 to +6.8 \\
RedimNet & +0.0012 & 0.0188 & -25.5 to +43.5 \\
Titanet & -0.0020 & 0.0146 & -13.5 to +15.3 \\
UniSpeech-SAT & -0.0216 & 0.0178 & -1.5 to +1.4 \\
WavLM-Base & -0.0126 & 0.0499 & -7.0 to +6.4 \\
WavLM-Base+ & -0.0352 & 0.0252 & -1.7 to +1.5 \\
WavLM-Large & -0.0004 & 0.0218 & -157.6 to +193.7 \\
X-Vector & +0.0014 & 0.0226 & -28.3 to +19.3 \\
\bottomrule%

%% file: generated/appendix/s2_per_encoder_ie.tex
E13 & CAM++ & SV & +0.0028 & 0.0103 & -0.0127 & +0.0248 & 1.226 & 12 \\
E13 & ECAPA-TDNN & SV & +0.0019 & 0.0141 & -0.0154 & +0.0218 & 1.020 & 12 \\
E13 & ECAPA-TDNN-Large & SV & -0.0050 & 0.0158 & -0.0467 & +0.0112 & 0.940 & 12 \\
E13 & HuBERT-Large & SSL & +0.0003 & 0.0334 & -0.0668 & +0.0520 & 1.007 & 12 \\
E13 & RawNet3 & SV & -0.0046 & 0.0178 & -0.0491 & +0.0240 & 0.949 & 12 \\
E13 & RedimNet & SV & +0.0087 & 0.0116 & -0.0119 & +0.0297 & 1.356 & 12 \\
E13 & Titanet & SV & +0.0005 & 0.0129 & -0.0151 & +0.0263 & 1.016 & 12 \\
E13 & UniSpeech-SAT & SSL & -0.0003 & 0.0154 & -0.0430 & +0.0126 & 1.000 & 12 \\
E13 & WavLM-Base & SSL & -0.0136 & 0.0210 & -0.0730 & +0.0020 & 0.931 & 12 \\
E13 & WavLM-Base+ & SSL & -0.0100 & 0.0131 & -0.0480 & +0.0033 & 0.962 & 12 \\
E13 & WavLM-Large & SSL & -0.0077 & 0.0149 & -0.0453 & +0.0060 & 0.965 & 12 \\
E13 & X-Vector & SV & +0.0140 & 0.0407 & -0.0231 & +0.1011 & 1.118 & 12 \\
E14 & CAM++ & SV & +0.0084 & 0.0225 & +0.0000 & +0.1076 & 1.275 & 63 \\
E14 & ECAPA-TDNN & SV & +0.0134 & 0.0252 & -0.0243 & +0.1250 & 1.355 & 63 \\
E14 & ECAPA-TDNN-Large & SV & +0.0119 & 0.0358 & -0.0069 & +0.1389 & 1.284 & 63 \\
E14 & HuBERT-Large & SSL & -0.0364 & 0.0445 & -0.1250 & +0.0799 & 0.927 & 63 \\
E14 & RawNet3 & SV & -0.0031 & 0.0331 & -0.0521 & +0.1076 & 0.564 & 63 \\
E14 & RedimNet & SV & +0.0161 & 0.0320 & -0.0069 & +0.1146 & 6.753 & 63 \\
E14 & Titanet & SV & +0.0138 & 0.0325 & +0.0000 & +0.1285 & 1.484 & 63 \\
E14 & UniSpeech-SAT & SSL & -0.0728 & 0.0531 & -0.1701 & +0.0347 & 0.860 & 63 \\
E14 & WavLM-Base & SSL & -0.0469 & 0.0551 & -0.1562 & +0.0833 & 0.954 & 63 \\
E14 & WavLM-Base+ & SSL & -0.0279 & 0.0502 & -0.1215 & +0.0972 & 0.962 & 63 \\
E14 & WavLM-Large & SSL & -0.0484 & 0.0572 & -0.1285 & +0.0799 & 0.920 & 63 \\
E14 & X-Vector & SV & +0.0014 & 0.0230 & -0.0590 & +0.0451 & 0.911 & 63 \\
E17 & CAM++ & SV & +0.0007 & 0.0047 & -0.0071 & +0.0096 & 1.034 & 12 \\
E17 & ECAPA-TDNN & SV & +0.0044 & 0.0059 & -0.0034 & +0.0133 & 1.204 & 12 \\
E17 & ECAPA-TDNN-Large & SV & +0.0071 & 0.0093 & -0.0103 & +0.0194 & 1.183 & 12 \\
E17 & HuBERT-Large & SSL & -0.0030 & 0.0224 & -0.0484 & +0.0284 & 0.995 & 12 \\
E17 & RawNet3 & SV & +0.0063 & 0.0093 & -0.0064 & +0.0309 & 1.296 & 12 \\
E17 & RedimNet & SV & +0.0020 & 0.0106 & -0.0129 & +0.0244 & 1.022 & 12 \\
E17 & Titanet & SV & +0.0019 & 0.0048 & -0.0055 & +0.0087 & 1.265 & 12 \\
E17 & UniSpeech-SAT & SSL & +0.0186 & 0.0156 & -0.0174 & +0.0362 & 1.049 & 12 \\
E17 & WavLM-Base & SSL & +0.0074 & 0.0200 & -0.0217 & +0.0299 & 1.019 & 12 \\
E17 & WavLM-Base+ & SSL & -0.0012 & 0.0201 & -0.0252 & +0.0365 & 0.997 & 12 \\
E17 & WavLM-Large & SSL & +0.0028 & 0.0155 & -0.0276 & +0.0286 & 1.009 & 12 \\
E17 & X-Vector & SV & -0.0055 & 0.0138 & -0.0255 & +0.0175 & 0.957 & 12 \\

%% file: generated/appendix/s3_settings.tex
E13 & \texttt{middle amr} & 0.153 & 0.149 & 0.0025 & 0.0053 \\
E13 & \texttt{middle g711} & 0.148 & 0.143 & 0.0016 & 0.0089 \\
E13 & \texttt{middle gsm} & 0.158 & 0.153 & 0.0047 & 0.0054 \\
E13 & \texttt{middle opus} & 0.134 & 0.137 & -0.0019 & -0.0055 \\
E13 & \texttt{older amr} & 0.141 & 0.146 & 0.0117 & -0.0265 \\
E13 & \texttt{older g711} & 0.135 & 0.140 & 0.0016 & -0.0132 \\
E13 & \texttt{older gsm} & 0.143 & 0.149 & 0.0220 & -0.0444 \\
E13 & \texttt{older opus} & 0.132 & 0.134 & -0.0048 & 0.0009 \\
E13 & \texttt{young amr} & 0.123 & 0.126 & -0.0063 & -0.0004 \\
E13 & \texttt{young g711} & 0.121 & 0.120 & 0.0003 & 0.0016 \\
E13 & \texttt{young gsm} & 0.127 & 0.130 & 0.0006 & -0.0065 \\
E13 & \texttt{young opus} & 0.114 & 0.115 & -0.0005 & -0.0006 \\
E14 & \texttt{ravdess angry crosstalk 15dB} & 0.267 & 0.269 & 0.0253 & -0.0396 \\
E14 & \texttt{ravdess angry crosstalk 25dB} & 0.214 & 0.229 & 0.0045 & -0.0410 \\
E14 & \texttt{ravdess angry crosstalk 5dB} & 0.380 & 0.367 & 0.0615 & -0.0563 \\
E14 & \texttt{ravdess angry environmental 15dB} & 0.182 & 0.202 & 0.0005 & -0.0500 \\
E14 & \texttt{ravdess angry environmental 25dB} & 0.160 & 0.194 & -0.0015 & -0.0792 \\
E14 & \texttt{ravdess angry environmental 5dB} & 0.196 & 0.219 & 0.0015 & -0.0590 \\
E14 & \texttt{ravdess angry gaussian 15dB} & 0.112 & 0.102 & -0.0050 & 0.0312 \\
E14 & \texttt{ravdess angry gaussian 25dB} & 0.094 & 0.096 & -0.0055 & 0.0028 \\
E14 & \texttt{ravdess angry gaussian 5dB} & 0.136 & 0.114 & 0.0005 & 0.0528 \\
E14 & \texttt{ravdess calm crosstalk 15dB} & 0.239 & 0.262 & -0.0050 & -0.0486 \\
E14 & \texttt{ravdess calm crosstalk 25dB} & 0.194 & 0.222 & -0.0010 & -0.0674 \\
E14 & \texttt{ravdess calm crosstalk 5dB} & 0.371 & 0.361 & 0.0437 & -0.0375 \\
E14 & \texttt{ravdess calm environmental 15dB} & 0.175 & 0.196 & -0.0010 & -0.0479 \\
E14 & \texttt{ravdess calm environmental 25dB} & 0.156 & 0.187 & -0.0045 & -0.0687 \\
E14 & \texttt{ravdess calm environmental 5dB} & 0.196 & 0.213 & 0.0020 & -0.0424 \\
E14 & \texttt{ravdess calm gaussian 15dB} & 0.097 & 0.096 & -0.0040 & 0.0076 \\
E14 & \texttt{ravdess calm gaussian 25dB} & 0.067 & 0.090 & -0.0040 & -0.0479 \\
E14 & \texttt{ravdess calm gaussian 5dB} & 0.098 & 0.108 & -0.0025 & -0.0187 \\
E14 & \texttt{ravdess disgust crosstalk 15dB} & 0.255 & 0.285 & 0.0263 & -0.1090 \\
E14 & \texttt{ravdess disgust crosstalk 25dB} & 0.209 & 0.245 & 0.0040 & -0.0917 \\
E14 & \texttt{ravdess disgust crosstalk 5dB} & 0.387 & 0.383 & 0.0685 & -0.0868 \\
E14 & \texttt{ravdess disgust environmental 15dB} & 0.178 & 0.218 & 0.0005 & -0.0979 \\
E14 & \texttt{ravdess disgust environmental 25dB} & 0.165 & 0.209 & -0.0035 & -0.1014 \\
E14 & \texttt{ravdess disgust environmental 5dB} & 0.204 & 0.235 & 0.0040 & -0.0812 \\
E14 & \texttt{ravdess disgust gaussian 15dB} & 0.122 & 0.118 & -0.0040 & 0.0153 \\
E14 & \texttt{ravdess disgust gaussian 25dB} & 0.124 & 0.112 & -0.0060 & 0.0361 \\
E14 & \texttt{ravdess disgust gaussian 5dB} & 0.120 & 0.130 & -0.0025 & -0.0215 \\
E14 & \texttt{ravdess fearful crosstalk 15dB} & 0.256 & 0.263 & 0.0322 & -0.0632 \\
E14 & \texttt{ravdess fearful crosstalk 25dB} & 0.213 & 0.223 & 0.0129 & -0.0431 \\
E14 & \texttt{ravdess fearful crosstalk 5dB} & 0.401 & 0.362 & 0.0908 & -0.0340 \\
E14 & \texttt{ravdess fearful environmental 15dB} & 0.172 & 0.197 & 0.0000 & -0.0590 \\
E14 & \texttt{ravdess fearful environmental 25dB} & 0.149 & 0.188 & -0.0010 & -0.0917 \\
E14 & \texttt{ravdess fearful environmental 5dB} & 0.194 & 0.214 & -0.0000 & -0.0479 \\
E14 & \texttt{ravdess fearful gaussian 15dB} & 0.108 & 0.097 & -0.0015 & 0.0299 \\
E14 & \texttt{ravdess fearful gaussian 25dB} & 0.095 & 0.091 & -0.0015 & 0.0132 \\
E14 & \texttt{ravdess fearful gaussian 5dB} & 0.119 & 0.109 & 0.0030 & 0.0215 \\
E14 & \texttt{ravdess happy crosstalk 15dB} & 0.247 & 0.280 & 0.0079 & -0.0889 \\
E14 & \texttt{ravdess happy crosstalk 25dB} & 0.216 & 0.240 & 0.0055 & -0.0653 \\
E14 & \texttt{ravdess happy crosstalk 5dB} & 0.369 & 0.378 & 0.0382 & -0.0757 \\
E14 & \texttt{ravdess happy environmental 15dB} & 0.177 & 0.213 & -0.0010 & -0.0868 \\
E14 & \texttt{ravdess happy environmental 25dB} & 0.159 & 0.205 & -0.0020 & -0.1069 \\
E14 & \texttt{ravdess happy environmental 5dB} & 0.201 & 0.230 & -0.0015 & -0.0674 \\
E14 & \texttt{ravdess happy gaussian 15dB} & 0.105 & 0.113 & -0.0045 & -0.0132 \\
E14 & \texttt{ravdess happy gaussian 25dB} & 0.096 & 0.107 & -0.0045 & -0.0201 \\
E14 & \texttt{ravdess happy gaussian 5dB} & 0.122 & 0.125 & -0.0020 & -0.0042 \\
E14 & \texttt{ravdess sad crosstalk 15dB} & 0.254 & 0.281 & 0.0099 & -0.0785 \\
E14 & \texttt{ravdess sad crosstalk 25dB} & 0.203 & 0.241 & 0.0010 & -0.0931 \\
E14 & \texttt{ravdess sad crosstalk 5dB} & 0.381 & 0.380 & 0.0506 & -0.0674 \\
E14 & \texttt{ravdess sad environmental 15dB} & 0.175 & 0.214 & -0.0055 & -0.0861 \\
E14 & \texttt{ravdess sad environmental 25dB} & 0.161 & 0.206 & -0.0050 & -0.1007 \\
E14 & \texttt{ravdess sad environmental 5dB} & 0.194 & 0.231 & -0.0045 & -0.0847 \\
E14 & \texttt{ravdess sad gaussian 15dB} & 0.115 & 0.114 & -0.0079 & 0.0139 \\
E14 & \texttt{ravdess sad gaussian 25dB} & 0.107 & 0.108 & -0.0079 & 0.0083 \\
E14 & \texttt{ravdess sad gaussian 5dB} & 0.114 & 0.126 & -0.0050 & -0.0215 \\
E14 & \texttt{ravdess surprise crosstalk 15dB} & 0.276 & 0.284 & 0.0536 & -0.0951 \\
E14 & \texttt{ravdess surprise crosstalk 25dB} & 0.229 & 0.244 & 0.0228 & -0.0674 \\
E14 & \texttt{ravdess surprise crosstalk 5dB} & 0.397 & 0.383 & 0.0873 & -0.0889 \\
E14 & \texttt{ravdess surprise environmental 15dB} & 0.188 & 0.218 & 0.0005 & -0.0708 \\
E14 & \texttt{ravdess surprise environmental 25dB} & 0.172 & 0.209 & -0.0020 & -0.0868 \\
E14 & \texttt{ravdess surprise environmental 5dB} & 0.205 & 0.235 & 0.0055 & -0.0785 \\
E14 & \texttt{ravdess surprise gaussian 15dB} & 0.120 & 0.117 & -0.0020 & 0.0090 \\
E14 & \texttt{ravdess surprise gaussian 25dB} & 0.115 & 0.111 & -0.0050 & 0.0146 \\
E14 & \texttt{ravdess surprise gaussian 5dB} & 0.132 & 0.129 & 0.0064 & -0.0035 \\
E17 & \texttt{1yr amr} & 0.202 & 0.193 & -0.0007 & 0.0235 \\
E17 & \texttt{1yr g711} & 0.196 & 0.187 & -0.0030 & 0.0254 \\
E17 & \texttt{1yr gsm} & 0.206 & 0.203 & -0.0031 & 0.0120 \\
E17 & \texttt{1yr opus} & 0.183 & 0.177 & -0.0053 & 0.0220 \\
E17 & \texttt{3yr amr} & 0.198 & 0.197 & 0.0098 & -0.0105 \\
E17 & \texttt{3yr g711} & 0.192 & 0.191 & 0.0052 & -0.0069 \\
E17 & \texttt{3yr gsm} & 0.209 & 0.207 & 0.0155 & -0.0160 \\
E17 & \texttt{3yr opus} & 0.178 & 0.181 & 0.0006 & -0.0089 \\
E17 & \texttt{5yr amr} & 0.199 & 0.193 & 0.0048 & 0.0075 \\
E17 & \texttt{5yr g711} & 0.190 & 0.187 & -0.0007 & 0.0075 \\
E17 & \texttt{5yr gsm} & 0.209 & 0.203 & 0.0089 & 0.0021 \\
E17 & \texttt{5yr opus} & 0.176 & 0.177 & -0.0029 & 0.0013 \\

%% file: generated/appendix/s4_agreement.tex
CAM++ & ECAPA-TDNN-Large & 339 & 20.9 & 0.993 & 0.0075 \\
CAM++ & Titanet & 339 & 20.6 & 0.984 & 0.0126 \\
ECAPA-TDNN-Large & Titanet & 339 & 19.5 & 0.982 & 0.0139 \\
ECAPA-TDNN-Large & RedimNet & 339 & 18.0 & 0.919 & 0.0184 \\
ECAPA-TDNN & Titanet & 339 & 16.5 & 0.846 & 0.0270 \\
CAM++ & RedimNet & 339 & 16.2 & 0.921 & 0.0187 \\
RedimNet & Titanet & 339 & 16.2 & 0.926 & 0.0192 \\
CAM++ & RawNet3 & 339 & 15.0 & 0.969 & 0.0191 \\
CAM++ & ECAPA-TDNN & 339 & 14.7 & 0.829 & 0.0296 \\
RawNet3 & Titanet & 339 & 14.7 & 0.973 & 0.0171 \\
ECAPA-TDNN & RedimNet & 339 & 14.2 & 0.812 & 0.0326 \\
ECAPA-TDNN & ECAPA-TDNN-Large & 339 & 13.9 & 0.816 & 0.0309 \\
\bottomrule%

%% file: generated/appendix/s1_conditions.tex
\Needspace{11\baselineskip}
\subsection{\texorpdfstring{E01 -- Emotional speech}{E01 -- Emotional speech}}
\label{sup:E01emotionalspeech}
{\footnotesize\setlength{\tabcolsep}{4pt}
\begin{longtable}{@{}p{0.26\textwidth}lrrrrrr@{}}
\caption{E01 -- Emotional speech: condition-level means over the contributing encoder rows. SD is over contributing encoder rows, not a confidence interval.}\label{sup:tab:E01emotionalspeech}\\
\toprule Condition & Corpus & Enc. & EER & SD & minDCF & AUC & Trials \\
\midrule
\endfirsthead
\multicolumn{8}{c}{\tablename~\thetable{} -- continued}\\
\toprule Condition & Corpus & Enc. & EER & SD & minDCF & AUC & Trials \\
\midrule
\endhead
\bottomrule
\endfoot
\texttt{ravdess\allowbreak/neutral\allowbreak\_neutral} & ravdess & 12 & 0.083 & 0.114 & 0.424 & 0.951 & 120,000 \\
\texttt{ravdess\allowbreak/calm\allowbreak\_calm} & ravdess & 12 & 0.106 & 0.119 & 0.378 & 0.936 & 15,000 \\
\texttt{ravdess\allowbreak/surprise\allowbreak\_surprise} & ravdess & 12 & 0.116 & 0.117 & 0.501 & 0.927 & 15,000 \\
\texttt{ravdess\allowbreak/disgust\allowbreak\_disgust} & ravdess & 12 & 0.131 & 0.127 & 0.513 & 0.918 & 15,000 \\
\texttt{ravdess\allowbreak/neutral\allowbreak\_calm} & ravdess & 12 & 0.136 & 0.137 & 0.527 & 0.905 & 15,000 \\
\texttt{ravdess\allowbreak/happy\allowbreak\_happy} & ravdess & 12 & 0.151 & 0.128 & 0.580 & 0.899 & 15,000 \\
\texttt{ravdess\allowbreak/neutral\allowbreak\_surprise} & ravdess & 12 & 0.170 & 0.148 & 0.709 & 0.879 & 15,000 \\
\texttt{ravdess\allowbreak/sad\allowbreak\_sad} & ravdess & 12 & 0.177 & 0.118 & 0.603 & 0.881 & 15,000 \\
\end{longtable}}

\Needspace{11\baselineskip}
\subsection{\texorpdfstring{E02 -- Dysarthric speech}{E02 -- Dysarthric speech}}
\label{sup:E02pathologicalspeech}
{\footnotesize\setlength{\tabcolsep}{4pt}
\begin{longtable}{@{}p{0.26\textwidth}lrrrrrr@{}}
\caption{E02 -- Dysarthric speech: condition-level means over the contributing encoder rows. SD is over contributing encoder rows, not a confidence interval.}\label{sup:tab:E02pathologicalspeech}\\
\toprule Condition & Corpus & Enc. & EER & SD & minDCF & AUC & Trials \\
\midrule
\endfirsthead
\multicolumn{8}{c}{\tablename~\thetable{} -- continued}\\
\toprule Condition & Corpus & Enc. & EER & SD & minDCF & AUC & Trials \\
\midrule
\endhead
\bottomrule
\endfoot
\texttt{TORGO\allowbreak/dysarthria\allowbreak\_moderate\allowbreak\_matched} & torgo & 12 & 0.302 & 0.117 & 0.941 & 0.755 & 37,404 \\
\texttt{TORGO\allowbreak/control\allowbreak\_matched} & torgo & 12 & 0.318 & 0.120 & 0.962 & 0.735 & 39,689 \\
\texttt{TORGO\allowbreak/dysarthria\allowbreak\_mild\allowbreak\_matched} & torgo & 12 & 0.326 & 0.105 & 0.978 & 0.730 & 38,154 \\
\texttt{TORGO\allowbreak/dysarthria\allowbreak\_severe\allowbreak\_matched} & torgo & 12 & 0.327 & 0.106 & 0.977 & 0.726 & 38,459 \\
\end{longtable}}

\Needspace{11\baselineskip}
\subsection{\texorpdfstring{E03 -- Speaking rate}{E03 -- Speaking rate}}
\label{sup:E03speakingrate}
{\footnotesize\setlength{\tabcolsep}{4pt}
\begin{longtable}{@{}p{0.26\textwidth}lrrrrrr@{}}
\caption{E03 -- Speaking rate: condition-level means over the contributing encoder rows. SD is over contributing encoder rows, not a confidence interval.}\label{sup:tab:E03speakingrate}\\
\toprule Condition & Corpus & Enc. & EER & SD & minDCF & AUC & Trials \\
\midrule
\endfirsthead
\multicolumn{8}{c}{\tablename~\thetable{} -- continued}\\
\toprule Condition & Corpus & Enc. & EER & SD & minDCF & AUC & Trials \\
\midrule
\endhead
\bottomrule
\endfoot
\texttt{common\allowbreak\_voice\allowbreak/clean} & CV & 12 & 0.108 & 0.118 & 0.454 & 0.933 & 120,000 \\
\texttt{common\allowbreak\_voice\allowbreak/slow\allowbreak\_slow} & CV & 12 & 0.167 & 0.077 & 0.767 & 0.899 & 120,000 \\
\texttt{common\allowbreak\_voice\allowbreak/fast\allowbreak\_fast} & CV & 12 & 0.192 & 0.070 & 0.798 & 0.881 & 120,000 \\
\texttt{ears\allowbreak/clean} & ears & 12 & 0.244 & 0.174 & 0.658 & 0.796 & 119,990 \\
\texttt{common\allowbreak\_voice\allowbreak/fast\allowbreak\_normal} & CV & 12 & 0.336 & 0.091 & 0.927 & 0.718 & 120,000 \\
\texttt{common\allowbreak\_voice\allowbreak/normal\allowbreak\_fast} & CV & 12 & 0.337 & 0.090 & 0.928 & 0.718 & 120,000 \\
\texttt{common\allowbreak\_voice\allowbreak/normal\allowbreak\_slow} & CV & 12 & 0.350 & 0.093 & 0.931 & 0.699 & 120,000 \\
\texttt{common\allowbreak\_voice\allowbreak/slow\allowbreak\_normal} & CV & 12 & 0.351 & 0.093 & 0.930 & 0.698 & 120,000 \\
\end{longtable}}

\Needspace{11\baselineskip}
\subsection{\texorpdfstring{E04 -- Recorded voice style}{E04 -- Recorded voice style}}
\label{sup:E04voicemode}
{\footnotesize\setlength{\tabcolsep}{4pt}
\begin{longtable}{@{}p{0.26\textwidth}lrrrrrr@{}}
\caption{E04 -- Recorded voice style: condition-level means over the contributing encoder rows. SD is over contributing encoder rows, not a confidence interval.}\label{sup:tab:E04voicemode}\\
\toprule Condition & Corpus & Enc. & EER & SD & minDCF & AUC & Trials \\
\midrule
\endfirsthead
\multicolumn{8}{c}{\tablename~\thetable{} -- continued}\\
\toprule Condition & Corpus & Enc. & EER & SD & minDCF & AUC & Trials \\
\midrule
\endhead
\bottomrule
\endfoot
\texttt{shout\allowbreak\_shout} & ears & 12 & 0.148 & 0.183 & 0.428 & 0.870 & 13,332 \\
\texttt{modal\allowbreak\_modal} & ears & 12 & 0.148 & 0.182 & 0.424 & 0.875 & 13,332 \\
\texttt{whisper\allowbreak\_whisper} & ears & 12 & 0.157 & 0.165 & 0.477 & 0.880 & 13,332 \\
\texttt{modal\allowbreak\_shout} & ears & 12 & 0.162 & 0.189 & 0.441 & 0.864 & 12,840 \\
\texttt{shout\allowbreak\_modal} & ears & 12 & 0.171 & 0.201 & 0.449 & 0.853 & 12,840 \\
\texttt{whisper\allowbreak\_modal} & ears & 12 & 0.209 & 0.194 & 0.626 & 0.827 & 12,840 \\
\texttt{modal\allowbreak\_whisper} & ears & 12 & 0.209 & 0.193 & 0.657 & 0.831 & 12,840 \\
\texttt{whisper\allowbreak\_shout} & ears & 12 & 0.212 & 0.191 & 0.632 & 0.826 & 12,840 \\
\texttt{shout\allowbreak\_whisper} & ears & 12 & 0.217 & 0.194 & 0.645 & 0.824 & 12,840 \\
\end{longtable}}

\Needspace{11\baselineskip}
\subsection{\texorpdfstring{E05 -- Codec round trips}{E05 -- Codec round trips}}
\label{sup:E05bluetoothcodecs}
{\footnotesize\setlength{\tabcolsep}{4pt}
\begin{longtable}{@{}p{0.26\textwidth}lrrrrrr@{}}
\caption{E05 -- Codec round trips: condition-level means over the contributing encoder rows. SD is over contributing encoder rows, not a confidence interval.}\label{sup:tab:E05bluetoothcodecs}\\
\toprule Condition & Corpus & Enc. & EER & SD & minDCF & AUC & Trials \\
\midrule
\endfirsthead
\multicolumn{8}{c}{\tablename~\thetable{} -- continued}\\
\toprule Condition & Corpus & Enc. & EER & SD & minDCF & AUC & Trials \\
\midrule
\endhead
\bottomrule
\endfoot
\texttt{aac\allowbreak\_high\allowbreak\_matched} & -- & 12 & 0.246 & 0.172 & 0.669 & 0.797 & 119,970 \\
\texttt{clean} & -- & 12 & 0.246 & 0.172 & 0.660 & 0.796 & 119,970 \\
\texttt{aac\allowbreak\_medium\allowbreak\_matched} & -- & 12 & 0.246 & 0.171 & 0.671 & 0.797 & 119,970 \\
\texttt{aptx\allowbreak\_high\allowbreak\_matched} & -- & 12 & 0.246 & 0.172 & 0.662 & 0.796 & 119,970 \\
\texttt{sbc\allowbreak\_high\allowbreak\_matched} & -- & 12 & 0.246 & 0.172 & 0.669 & 0.796 & 119,970 \\
\texttt{aptx\allowbreak\_vs\allowbreak\_aac\allowbreak\_high} & -- & 12 & 0.246 & 0.171 & 0.671 & 0.796 & 119,970 \\
\texttt{aac\allowbreak\_vs\allowbreak\_aptx\allowbreak\_medium} & -- & 12 & 0.246 & 0.173 & 0.666 & 0.796 & 119,970 \\
\texttt{aptx\allowbreak\_low\allowbreak\_matched} & -- & 12 & 0.248 & 0.170 & 0.661 & 0.795 & 119,970 \\
\end{longtable}}

\Needspace{11\baselineskip}
\subsection{\texorpdfstring{E06 -- Packet loss}{E06 -- Packet loss}}
\label{sup:E06packetloss}
{\footnotesize\setlength{\tabcolsep}{4pt}
\begin{longtable}{@{}p{0.26\textwidth}lrrrrrr@{}}
\caption{E06 -- Packet loss: condition-level means over the contributing encoder rows. SD is over contributing encoder rows, not a confidence interval.}\label{sup:tab:E06packetloss}\\
\toprule Condition & Corpus & Enc. & EER & SD & minDCF & AUC & Trials \\
\midrule
\endfirsthead
\multicolumn{8}{c}{\tablename~\thetable{} -- continued}\\
\toprule Condition & Corpus & Enc. & EER & SD & minDCF & AUC & Trials \\
\midrule
\endhead
\bottomrule
\endfoot
\texttt{common\allowbreak\_voice\allowbreak/clean} & CV & 12 & 0.108 & 0.117 & 0.463 & 0.933 & 120,000 \\
\texttt{common\allowbreak\_voice\allowbreak/clean\allowbreak\_x\allowbreak\_0.5pct} & CV & 12 & 0.109 & 0.118 & 0.462 & 0.931 & 120,000 \\
\texttt{common\allowbreak\_voice\allowbreak/0.5pct\allowbreak\_0.5pct} & CV & 12 & 0.110 & 0.118 & 0.463 & 0.930 & 120,000 \\
\texttt{common\allowbreak\_voice\allowbreak/clean\allowbreak\_x\allowbreak\_2.0pct} & CV & 12 & 0.112 & 0.121 & 0.473 & 0.929 & 120,000 \\
\texttt{common\allowbreak\_voice\allowbreak/2.0pct\allowbreak\_2.0pct} & CV & 12 & 0.112 & 0.121 & 0.474 & 0.928 & 120,000 \\
\texttt{common\allowbreak\_voice\allowbreak/5.0pct\allowbreak\_5.0pct} & CV & 12 & 0.116 & 0.123 & 0.482 & 0.925 & 120,000 \\
\texttt{common\allowbreak\_voice\allowbreak/clean\allowbreak\_x\allowbreak\_5.0pct} & CV & 12 & 0.116 & 0.124 & 0.478 & 0.924 & 120,000 \\
\texttt{common\allowbreak\_voice\allowbreak/10.0pct\allowbreak\_10.0pct} & CV & 12 & 0.120 & 0.125 & 0.495 & 0.922 & 120,000 \\
\end{longtable}}

\Needspace{11\baselineskip}
\subsection{\texorpdfstring{E07 -- Temporal drift}{E07 -- Temporal drift}}
\label{sup:E07temporaldrift}
{\footnotesize\setlength{\tabcolsep}{4pt}
\begin{longtable}{@{}p{0.26\textwidth}lrrrrrr@{}}
\caption{E07 -- Temporal drift: condition-level means over the contributing encoder rows. SD is over contributing encoder rows, not a confidence interval.}\label{sup:tab:E07temporaldrift}\\
\toprule Condition & Corpus & Enc. & EER & SD & minDCF & AUC & Trials \\
\midrule
\endfirsthead
\multicolumn{8}{c}{\tablename~\thetable{} -- continued}\\
\toprule Condition & Corpus & Enc. & EER & SD & minDCF & AUC & Trials \\
\midrule
\endhead
\bottomrule
\endfoot
\texttt{voxceleb1\allowbreak/gap\allowbreak\_0yr} & voxceleb1 & 12 & 0.168 & 0.184 & 0.501 & 0.862 & 120,000 \\
\texttt{voxceleb1\allowbreak/gap\allowbreak\_5yr} & voxceleb1 & 12 & 0.176 & 0.188 & 0.542 & 0.855 & 120,000 \\
\texttt{voxceleb1\allowbreak/gap\allowbreak\_1yr} & voxceleb1 & 12 & 0.177 & 0.190 & 0.522 & 0.852 & 120,000 \\
\texttt{voxceleb1\allowbreak/gap\allowbreak\_3yr} & voxceleb1 & 12 & 0.180 & 0.193 & 0.533 & 0.848 & 120,000 \\
\end{longtable}}

\Needspace{11\baselineskip}
\subsection{\texorpdfstring{E08 -- L2 accent}{E08 -- L2 accent}}
\label{sup:E08accentstrength}
{\footnotesize\setlength{\tabcolsep}{4pt}
\begin{longtable}{@{}p{0.26\textwidth}lrrrrrr@{}}
\caption{E08 -- L2 accent: condition-level means over the contributing encoder rows. SD is over contributing encoder rows, not a confidence interval.}\label{sup:tab:E08accentstrength}\\
\toprule Condition & Corpus & Enc. & EER & SD & minDCF & AUC & Trials \\
\midrule
\endfirsthead
\multicolumn{8}{c}{\tablename~\thetable{} -- continued}\\
\toprule Condition & Corpus & Enc. & EER & SD & minDCF & AUC & Trials \\
\midrule
\endhead
\bottomrule
\endfoot
\texttt{l2\allowbreak\_arctic\allowbreak/level\allowbreak\_3\allowbreak\_vs\allowbreak\_native} & L2-ARCTIC/CV & 12 & 0.077 & 0.103 & 0.372 & 0.956 & 19,944 \\
\texttt{l2\allowbreak\_arctic\allowbreak/level\allowbreak\_1\allowbreak\_vs\allowbreak\_native} & L2-ARCTIC/CV & 12 & 0.079 & 0.105 & 0.386 & 0.954 & 19,908 \\
\texttt{l2\allowbreak\_arctic\allowbreak/level\allowbreak\_4\allowbreak\_vs\allowbreak\_native} & L2-ARCTIC/CV & 12 & 0.082 & 0.117 & 0.350 & 0.948 & 19,968 \\
\texttt{l2\allowbreak\_arctic\allowbreak/l2\allowbreak\_all\allowbreak\_vs\allowbreak\_native} & L2-ARCTIC/CV & 12 & 0.084 & 0.116 & 0.371 & 0.951 & 19,932 \\
\texttt{l2\allowbreak\_arctic\allowbreak/level\allowbreak\_2\allowbreak\_vs\allowbreak\_native} & L2-ARCTIC/CV & 12 & 0.088 & 0.116 & 0.376 & 0.944 & 19,932 \\
\texttt{l2\allowbreak\_arctic\allowbreak/level\allowbreak\_5\allowbreak\_vs\allowbreak\_native} & L2-ARCTIC/CV & 12 & 0.091 & 0.120 & 0.388 & 0.944 & 19,932 \\
\texttt{cv\allowbreak\_en\allowbreak/native\allowbreak\_matched} & CV & 12 & 0.110 & 0.118 & 0.411 & 0.935 & 19,992 \\
\texttt{l2\allowbreak\_arctic\allowbreak/level\allowbreak\_3\allowbreak\_matched} & L2-ARCTIC & 12 & 0.142 & 0.171 & 0.430 & 0.888 & 23,892 \\
\end{longtable}}

\Needspace{11\baselineskip}
\subsection{\texorpdfstring{E10 -- TTS spoofing}{E10 -- TTS spoofing}}
\label{sup:E10diffusiontts}
{\footnotesize\setlength{\tabcolsep}{4pt}
\begin{longtable}{@{}p{0.26\textwidth}lrrrrrr@{}}
\caption{E10 -- TTS spoofing: condition-level means over the contributing encoder rows. SD is over contributing encoder rows, not a confidence interval.}\label{sup:tab:E10diffusiontts}\\
\toprule Condition & Corpus & Enc. & EER & SD & minDCF & AUC & Trials \\
\midrule
\endfirsthead
\multicolumn{8}{c}{\tablename~\thetable{} -- continued}\\
\toprule Condition & Corpus & Enc. & EER & SD & minDCF & AUC & Trials \\
\midrule
\endhead
\bottomrule
\endfoot
\texttt{real\allowbreak\_real\allowbreak\_baseline} & -- & 12 & 0.239 & 0.170 & 0.649 & 0.804 & 120,000 \\
\texttt{voicecraft\allowbreak/cross\allowbreak\_utt\allowbreak\_spoof} & -- & 12 & 0.311 & 0.094 & 0.781 & 0.750 & 113,664 \\
\texttt{e2\allowbreak\_tts\allowbreak/cross\allowbreak\_utt\allowbreak\_spoof} & -- & 12 & 0.368 & 0.050 & 0.913 & 0.691 & 118,104 \\
\texttt{voicecraft\allowbreak/synth\allowbreak\_vs\allowbreak\_real} & -- & 12 & 0.433 & 0.127 & 0.904 & 0.584 & 113,664 \\
\texttt{voicecraft\allowbreak/real\allowbreak\_vs\allowbreak\_synth} & -- & 12 & 0.436 & 0.130 & 0.901 & 0.581 & 113,568 \\
\texttt{e2\allowbreak\_tts\allowbreak/synth\allowbreak\_vs\allowbreak\_real} & -- & 12 & 0.530 & 0.080 & 0.996 & 0.465 & 118,104 \\
\texttt{e2\allowbreak\_tts\allowbreak/real\allowbreak\_vs\allowbreak\_synth} & -- & 12 & 0.530 & 0.079 & 0.992 & 0.466 & 118,344 \\
\end{longtable}}

\Needspace{11\baselineskip}
\subsection{\texorpdfstring{E11 -- Voice conversion}{E11 -- Voice conversion}}
\label{sup:E11realtimevc}
{\footnotesize\setlength{\tabcolsep}{4pt}
\begin{longtable}{@{}p{0.26\textwidth}lrrrrrr@{}}
\caption{E11 -- Voice conversion: condition-level means over the contributing encoder rows. SD is over contributing encoder rows, not a confidence interval.}\label{sup:tab:E11realtimevc}\\
\toprule Condition & Corpus & Enc. & EER & SD & minDCF & AUC & Trials \\
\midrule
\endfirsthead
\multicolumn{8}{c}{\tablename~\thetable{} -- continued}\\
\toprule Condition & Corpus & Enc. & EER & SD & minDCF & AUC & Trials \\
\midrule
\endhead
\bottomrule
\endfoot
\texttt{freevc\allowbreak/spoof\allowbreak\_vs\allowbreak\_genuine} & -- & 12 & 0.225 & 0.096 & 0.620 & 0.832 & 60,000 \\
\texttt{real\allowbreak\_real\allowbreak\_baseline} & -- & 12 & 0.241 & 0.166 & 0.700 & 0.801 & 60,000 \\
\texttt{knn\allowbreak\_vc\allowbreak/spoof\allowbreak\_vs\allowbreak\_genuine} & -- & 12 & 0.357 & 0.060 & 0.890 & 0.690 & 60,000 \\
\end{longtable}}

\Needspace{11\baselineskip}
\subsection{\texorpdfstring{E13 -- Age $\times$ codec}{E13 -- Age x codec}}
\label{sup:E13pairedagexcodec}
{\footnotesize\setlength{\tabcolsep}{4pt}
\begin{longtable}{@{}p{0.26\textwidth}lrrrrrr@{}}
\caption{E13 -- Age $\times$ codec: condition-level means over the contributing encoder rows. SD is over contributing encoder rows, not a confidence interval.}\label{sup:tab:E13pairedagexcodec}\\
\toprule Condition & Corpus & Enc. & EER & SD & minDCF & AUC & Trials \\
\midrule
\endfirsthead
\multicolumn{8}{c}{\tablename~\thetable{} -- continued}\\
\toprule Condition & Corpus & Enc. & EER & SD & minDCF & AUC & Trials \\
\midrule
\endhead
\bottomrule
\endfoot
\texttt{all\allowbreak\_\allowbreak\_clean} & CV & 12 & 0.108 & 0.119 & 0.462 & 0.933 & 120,000 \\
\texttt{young\allowbreak\_\allowbreak\_clean} & CV & 12 & 0.111 & 0.117 & 0.449 & 0.929 & 39,996 \\
\texttt{all\allowbreak\_\allowbreak\_opus} & CV & 12 & 0.112 & 0.118 & 0.471 & 0.931 & 120,000 \\
\texttt{young\allowbreak\_\allowbreak\_opus} & CV & 12 & 0.114 & 0.116 & 0.462 & 0.928 & 39,996 \\
\texttt{all\allowbreak\_\allowbreak\_g711} & CV & 12 & 0.117 & 0.111 & 0.528 & 0.929 & 120,000 \\
\texttt{young\allowbreak\_\allowbreak\_g711} & CV & 12 & 0.121 & 0.111 & 0.498 & 0.924 & 39,996 \\
\texttt{young\allowbreak\_\allowbreak\_amr} & CV & 12 & 0.123 & 0.109 & 0.537 & 0.924 & 39,996 \\
\texttt{all\allowbreak\_\allowbreak\_amr} & CV & 12 & 0.124 & 0.108 & 0.555 & 0.927 & 120,000 \\
\end{longtable}}

\Needspace{11\baselineskip}
\subsection{\texorpdfstring{E14 -- Emotion $\times$ noise}{E14 -- Emotion x noise}}
\label{sup:E14matchedravdess}
{\footnotesize\setlength{\tabcolsep}{4pt}
\begin{longtable}{@{}p{0.26\textwidth}lrrrrrr@{}}
\caption{E14 -- Emotion $\times$ noise: condition-level means over the contributing encoder rows. SD is over contributing encoder rows, not a confidence interval.}\label{sup:tab:E14matchedravdess}\\
\toprule Condition & Corpus & Enc. & EER & SD & minDCF & AUC & Trials \\
\midrule
\endfirsthead
\multicolumn{8}{c}{\tablename~\thetable{} -- continued}\\
\toprule Condition & Corpus & Enc. & EER & SD & minDCF & AUC & Trials \\
\midrule
\endhead
\bottomrule
\endfoot
\texttt{neutral\allowbreak\_\allowbreak\_clean\allowbreak\_\allowbreak\_None} & ravdess & 12 & 0.054 & 0.079 & 0.168 & 0.972 & 3,456 \\
\texttt{calm\allowbreak\_\allowbreak\_gaussian\allowbreak\_\allowbreak\_25} & ravdess & 12 & 0.067 & 0.099 & 0.236 & 0.955 & 3,456 \\
\texttt{calm\allowbreak\_\allowbreak\_clean\allowbreak\_\allowbreak\_None} & ravdess & 12 & 0.072 & 0.088 & 0.226 & 0.964 & 3,456 \\
\texttt{neutral\allowbreak\_\allowbreak\_gaussian\allowbreak\_\allowbreak\_25} & ravdess & 12 & 0.072 & 0.107 & 0.257 & 0.953 & 3,456 \\
\texttt{fearful\allowbreak\_\allowbreak\_clean\allowbreak\_\allowbreak\_None} & ravdess & 12 & 0.073 & 0.092 & 0.228 & 0.962 & 3,456 \\
\texttt{angry\allowbreak\_\allowbreak\_clean\allowbreak\_\allowbreak\_None} & ravdess & 12 & 0.078 & 0.095 & 0.260 & 0.948 & 3,456 \\
\texttt{neutral\allowbreak\_\allowbreak\_gaussian\allowbreak\_\allowbreak\_15} & ravdess & 12 & 0.078 & 0.120 & 0.247 & 0.946 & 3,456 \\
\texttt{happy\allowbreak\_\allowbreak\_clean\allowbreak\_\allowbreak\_None} & ravdess & 12 & 0.089 & 0.108 & 0.263 & 0.942 & 3,456 \\
\end{longtable}}

\Needspace{11\baselineskip}
\subsection{\texorpdfstring{E15 -- Pathology and codec (descriptive)}{E15 -- Pathology and codec (descriptive)}}
\label{sup:E15pathologyxcodec}
{\footnotesize\setlength{\tabcolsep}{4pt}
\begin{longtable}{@{}p{0.26\textwidth}lrrrrrr@{}}
\caption{E15 -- Pathology and codec (descriptive): condition-level means over the contributing encoder rows. SD is over contributing encoder rows, not a confidence interval.}\label{sup:tab:E15pathologyxcodec}\\
\toprule Condition & Corpus & Enc. & EER & SD & minDCF & AUC & Trials \\
\midrule
\endfirsthead
\multicolumn{8}{c}{\tablename~\thetable{} -- continued}\\
\toprule Condition & Corpus & Enc. & EER & SD & minDCF & AUC & Trials \\
\midrule
\endhead
\bottomrule
\endfoot
\texttt{EARS\allowbreak/healthy\allowbreak\_clean} & ears & 12 & 0.246 & 0.175 & 0.663 & 0.797 & 119,988 \\
\texttt{EARS\allowbreak/healthy\allowbreak\_opus} & ears & 12 & 0.259 & 0.160 & 0.679 & 0.787 & 29,997 \\
\texttt{EARS\allowbreak/healthy\allowbreak\_amr} & ears & 12 & 0.260 & 0.165 & 0.702 & 0.786 & 29,996 \\
\texttt{EARS\allowbreak/healthy\allowbreak\_gsm} & ears & 12 & 0.260 & 0.163 & 0.718 & 0.788 & 29,996 \\
\texttt{EARS\allowbreak/healthy\allowbreak\_g711} & ears & 12 & 0.279 & 0.155 & 0.755 & 0.772 & 29,998 \\
\texttt{TORGO\allowbreak/dysarthria\allowbreak\_moderate\allowbreak\_clean} & torgo & 12 & 0.303 & 0.106 & 0.932 & 0.752 & 37,411 \\
\texttt{TORGO\allowbreak/dysarthria\allowbreak\_moderate\allowbreak\_g711} & torgo & 12 & 0.304 & 0.100 & 0.880 & 0.758 & 9,376 \\
\texttt{TORGO\allowbreak/dysarthria\allowbreak\_moderate\allowbreak\_gsm} & torgo & 12 & 0.305 & 0.113 & 0.921 & 0.744 & 9,367 \\
\end{longtable}}

\Needspace{11\baselineskip}
\subsection{\texorpdfstring{E16 -- Gender $\times$ noise}{E16 -- Gender x noise}}
\label{sup:E16genderxnoise}
{\footnotesize\setlength{\tabcolsep}{4pt}
\begin{longtable}{@{}p{0.26\textwidth}lrrrrrr@{}}
\caption{E16 -- Gender $\times$ noise: condition-level means over the contributing encoder rows. SD is over contributing encoder rows, not a confidence interval.}\label{sup:tab:E16genderxnoise}\\
\toprule Condition & Corpus & Enc. & EER & SD & minDCF & AUC & Trials \\
\midrule
\endfirsthead
\multicolumn{8}{c}{\tablename~\thetable{} -- continued}\\
\toprule Condition & Corpus & Enc. & EER & SD & minDCF & AUC & Trials \\
\midrule
\endhead
\bottomrule
\endfoot
\texttt{CV\allowbreak/all\allowbreak\_clean} & CV & 12 & 0.109 & 0.118 & 0.446 & 0.932 & 60,000 \\
\texttt{CV\allowbreak/F\allowbreak\_clean} & CV & 12 & 0.110 & 0.113 & 0.432 & 0.932 & 60,000 \\
\texttt{CV\allowbreak/M\allowbreak\_clean} & CV & 12 & 0.117 & 0.116 & 0.466 & 0.927 & 60,000 \\
\texttt{CV\allowbreak/all\allowbreak\_gaussian\allowbreak\_25dB} & CV & 12 & 0.131 & 0.135 & 0.453 & 0.912 & 13,332 \\
\texttt{CV\allowbreak/F\allowbreak\_environmental\allowbreak\_25dB} & CV & 12 & 0.131 & 0.138 & 0.448 & 0.910 & 6,660 \\
\texttt{CV\allowbreak/all\allowbreak\_gaussian\allowbreak\_15dB} & CV & 12 & 0.137 & 0.143 & 0.496 & 0.900 & 13,332 \\
\texttt{CV\allowbreak/F\allowbreak\_gaussian\allowbreak\_25dB} & CV & 12 & 0.139 & 0.134 & 0.450 & 0.910 & 6,660 \\
\texttt{CV\allowbreak/F\allowbreak\_gaussian\allowbreak\_15dB} & CV & 12 & 0.139 & 0.137 & 0.503 & 0.906 & 6,660 \\
\end{longtable}}

\Needspace{11\baselineskip}
\subsection{\texorpdfstring{E17 -- Temporal drift $\times$ codec}{E17 -- Temporal drift x codec}}
\label{sup:E17temporaldriftxcodec}
{\footnotesize\setlength{\tabcolsep}{4pt}
\begin{longtable}{@{}p{0.26\textwidth}lrrrrrr@{}}
\caption{E17 -- Temporal drift $\times$ codec: condition-level means over the contributing encoder rows. SD is over contributing encoder rows, not a confidence interval.}\label{sup:tab:E17temporaldriftxcodec}\\
\toprule Condition & Corpus & Enc. & EER & SD & minDCF & AUC & Trials \\
\midrule
\endfirsthead
\multicolumn{8}{c}{\tablename~\thetable{} -- continued}\\
\toprule Condition & Corpus & Enc. & EER & SD & minDCF & AUC & Trials \\
\midrule
\endhead
\bottomrule
\endfoot
\texttt{VoxCeleb1\allowbreak/gap\allowbreak\_0\allowbreak\_clean} & voxceleb1 & 12 & 0.168 & 0.186 & 0.502 & 0.862 & 120,000 \\
\texttt{VoxCeleb1\allowbreak/gap\allowbreak\_0\allowbreak\_opus} & voxceleb1 & 12 & 0.169 & 0.180 & 0.496 & 0.862 & 30,000 \\
\texttt{VoxCeleb1\allowbreak/gap\allowbreak\_5yr\allowbreak\_opus} & voxceleb1 & 12 & 0.176 & 0.183 & 0.523 & 0.860 & 9,984 \\
\texttt{VoxCeleb1\allowbreak/gap\allowbreak\_1yr\allowbreak\_clean} & voxceleb1 & 12 & 0.176 & 0.189 & 0.516 & 0.853 & 39,984 \\
\texttt{VoxCeleb1\allowbreak/gap\allowbreak\_5yr\allowbreak\_clean} & voxceleb1 & 12 & 0.176 & 0.187 & 0.516 & 0.856 & 39,984 \\
\texttt{VoxCeleb1\allowbreak/gap\allowbreak\_3yr\allowbreak\_opus} & voxceleb1 & 12 & 0.178 & 0.181 & 0.496 & 0.854 & 9,984 \\
\texttt{VoxCeleb1\allowbreak/gap\allowbreak\_0\allowbreak\_g711} & voxceleb1 & 12 & 0.179 & 0.177 & 0.567 & 0.856 & 30,000 \\
\texttt{VoxCeleb1\allowbreak/gap\allowbreak\_3yr\allowbreak\_clean} & voxceleb1 & 12 & 0.180 & 0.192 & 0.527 & 0.849 & 39,984 \\
\end{longtable}}

\Needspace{11\baselineskip}
\subsection{\texorpdfstring{E18 -- Accent and codec (descriptive)}{E18 -- Accent and codec (descriptive)}}
\label{sup:E18accentxcodec}
{\footnotesize\setlength{\tabcolsep}{4pt}
\begin{longtable}{@{}p{0.26\textwidth}lrrrrrr@{}}
\caption{E18 -- Accent and codec (descriptive): condition-level means over the contributing encoder rows. SD is over contributing encoder rows, not a confidence interval.}\label{sup:tab:E18accentxcodec}\\
\toprule Condition & Corpus & Enc. & EER & SD & minDCF & AUC & Trials \\
\midrule
\endfirsthead
\multicolumn{8}{c}{\tablename~\thetable{} -- continued}\\
\toprule Condition & Corpus & Enc. & EER & SD & minDCF & AUC & Trials \\
\midrule
\endhead
\bottomrule
\endfoot
\texttt{CV\allowbreak\_EN\allowbreak/native\allowbreak\_gsm} & CV & 12 & 0.105 & 0.117 & 0.388 & 0.936 & 15,000 \\
\texttt{CV\allowbreak\_EN\allowbreak/native\allowbreak\_amr} & CV & 12 & 0.106 & 0.119 & 0.381 & 0.934 & 15,000 \\
\texttt{CV\allowbreak\_EN\allowbreak/native\allowbreak\_clean} & CV & 12 & 0.109 & 0.119 & 0.438 & 0.933 & 60,000 \\
\texttt{CV\allowbreak\_EN\allowbreak/native\allowbreak\_opus} & CV & 12 & 0.113 & 0.120 & 0.424 & 0.930 & 15,000 \\
\texttt{CV\allowbreak\_EN\allowbreak/native\allowbreak\_g711} & CV & 12 & 0.115 & 0.111 & 0.456 & 0.929 & 15,000 \\
\texttt{L2-ARCTIC\allowbreak/accent\allowbreak\_gsm} & L2-ARCTIC & 12 & 0.175 & 0.171 & 0.649 & 0.859 & 14,951 \\
\texttt{L2-ARCTIC\allowbreak/accent\allowbreak\_amr} & L2-ARCTIC & 12 & 0.177 & 0.174 & 0.551 & 0.854 & 14,951 \\
\texttt{L2-ARCTIC\allowbreak/accent\allowbreak\_clean} & L2-ARCTIC & 12 & 0.179 & 0.169 & 0.650 & 0.856 & 59,828 \\
\texttt{L2-ARCTIC\allowbreak/accent\allowbreak\_opus} & L2-ARCTIC & 12 & 0.180 & 0.166 & 0.673 & 0.853 & 14,957 \\
\texttt{L2-ARCTIC\allowbreak/accent\allowbreak\_g711} & L2-ARCTIC & 12 & 0.188 & 0.168 & 0.682 & 0.850 & 14,961 \\
\end{longtable}}

\Needspace{11\baselineskip}
\subsection{\texorpdfstring{E20 -- Age $\times$ codec $\times$ drift (age proxied)}{E20 -- Age x codec x drift (age proxied)}}
\label{sup:E20agexcodecxdrift}
{\footnotesize\renewcommand{\arraystretch}{0.92}\setlength{\tabcolsep}{4pt}
\begin{longtable}{@{}p{0.26\textwidth}lrrrrrr@{}}
\caption{E20 -- Age $\times$ codec $\times$ drift (age proxied): condition-level means over the contributing encoder rows. SD is over contributing encoder rows, not a confidence interval.}\label{sup:tab:E20agexcodecxdrift}\\
\toprule Condition & Corpus & Enc. & EER & SD & minDCF & AUC & Trials \\
\midrule
\endfirsthead
\multicolumn{8}{c}{\tablename~\thetable{} -- continued}\\
\toprule Condition & Corpus & Enc. & EER & SD & minDCF & AUC & Trials \\
\midrule
\endhead
\bottomrule
\endfoot
\texttt{VoxCeleb1\allowbreak/age\allowbreak\_middle\allowbreak\_clean} & voxceleb1 & 12 & 0.163 & 0.180 & 0.463 & 0.868 & 15,000 \\
\texttt{VoxCeleb1\allowbreak/AC\allowbreak\_older\allowbreak\_opus} & voxceleb1 & 12 & 0.163 & 0.178 & 0.488 & 0.869 & 7,500 \\
\texttt{VoxCeleb1\allowbreak\_3way\allowbreak/older\allowbreak\_gap5yr\allowbreak\_opus} & voxceleb1 & 12 & 0.166 & 0.178 & 0.491 & 0.862 & 3,312 \\
\texttt{VoxCeleb1\allowbreak/gap0\allowbreak\_opus} & voxceleb1 & 12 & 0.167 & 0.180 & 0.471 & 0.867 & 9,996 \\
\texttt{VoxCeleb1\allowbreak/AC\allowbreak\_middle\allowbreak\_opus} & voxceleb1 & 12 & 0.167 & 0.177 & 0.469 & 0.866 & 7,500 \\
\texttt{VoxCeleb1\allowbreak/age\allowbreak\_older\allowbreak\_clean} & voxceleb1 & 12 & 0.167 & 0.183 & 0.473 & 0.865 & 15,000 \\
\texttt{VoxCeleb1\allowbreak/clean} & voxceleb1 & 12 & 0.168 & 0.184 & 0.488 & 0.863 & 30,000 \\
\texttt{VoxCeleb1\allowbreak/BC\allowbreak\_older\allowbreak\_gap5yr} & voxceleb1 & 12 & 0.168 & 0.181 & 0.476 & 0.864 & 7,488 \\
\end{longtable}}